\documentclass{aa}  
\usepackage{graphicx}
\usepackage{txfonts}
\usepackage{natbib}
\pdfoutput=1
\bibpunct{(}{)}{;}{a}{}{,}   
\def\lsim{\mathrel{\rlap{\lower 3pt \hbox{$\sim$}} \raise 2.0pt \hbox{$<$}}}
\def\gsim{\mathrel{\rlap{\lower 3pt \hbox{$\sim$}} \raise 2.0pt \hbox{$>$}}}

\newcommand{\msun}{\ensuremath{M_\odot}}

\begin{document}
\title{A panoramic VISTA of the stellar halo of  NGC 253 \thanks{Based on observations taken within the VISTA Science
    Verification Program ID 60.A-9285(A)}}

%
   
\author{L.~Greggio\inst{1}
       \and M.~Rejkuba\inst{2,3}
       \and O.~A. Gonzalez\inst{4}
       \and M.~Arnaboldi\inst{2}
       \and E.~Iodice\inst{5}
       \and M.~Irwin\inst{6}
       \and M.J.~Neeser\inst{2}
       \and J.~Emerson\inst{7}
}

   \offprints{L. Greggio}

   \institute{
               INAF, Osservatorio Astronomico di Padova, Vicolo dell'Osservatorio 5,
35122 Padova, Italy\\
             \email{greggio@pd.astro.it}
        \and
          ESO, Karl-Schwarzschild-Strasse 2, D-85748 Garching, Germany\\
              \email{[mrejkuba],[marnabol],[mneeser]@eso.org}
        \and
          Excellence Cluster Universe, Boltzmannstr.\ 2, D-85748, Garching, Germany 
        \and
          ESO, Ave. Alonso de Cordova 3107, Casilla 19, Santiago 19001, Chile\\
              \email{ogonzale@eso.org}
        \and
          INAF, Osservatorio Astronomico di Capodimonte, 80126, Napoli, Italy
        \and
          Institute of Astronomy, Madingley Road, Cambridge CB03 0HA, UK
        \and
          Astronomy Unit, School of Physics and Astronomy, Queen Mary University of London, Mile End Road, London, E1 4NS, UK
             }

   \date{Draft date 12.09.2013; Received date; accepted date}
\titlerunning{Stellar halo of NGC~253}

  \abstract
   {Outskirts of large galaxies contain important information about galaxy formation and assembly. Resolved star count studies can probe the extremely low surface brightness of the outer halos.}
   {NGC 253 is a nearly edge-on disk galaxy in the Sculptor group
     where we resolved the halo stars from ground-based images, with
     the aim of studying its stellar population content, the structure
     and the overall extent of the halo. }
   {We use Z and J-band images from the VIRCAM camera mounted on the
     VISTA telescope to construct the spatially resolved J vs. Z-J
     colour-magnitude diagrams (CMDs).  The very deep photometry and the wide
     area covered allows us to trace the red giant branch (RGB) and asymptotic giant branch (AGB) stars that belong to the halo of NGC 253 out to 50 kpc along the galaxy minor axis.}
   {We confirm the existence of an extra planar stellar component of
     the disk, with a very prominent southern shelf and a symmetrical
     feature on the north side. The only additional visible sub-structure 
     is an overdensity in the north-west part of the halo $\sim 28$~kpc distant from the
     plane and extending over  $20$~kpc  parallel with the disk. We measure the transition from the disk to
     the halo at a radial distance of about 25 kpc with a clear break
     in the number density profile. The isodensity contours show that
     the inner halo is a flattened structure that blends
     with a more extended, diffuse, rounder outer halo. Such external
     structure can be traced to the very edge of our image out to 50
     kpc from the disk plane. The number density profile of the stars in the
     stellar halo follows a power law with index $-1.6$, as function
     of radius.  The CMD shows a very homogeneous stellar population across the
     field; by comparing with isochrones  we conclude that the
     RGB stars are $\sim 8$ Gyr old or more, 
     while the AGB stars trace a population of about $2 \times 10^8$
     M$_\odot$ formed from $\sim 0.5$ to a few  Gyr ago. Surprisingly, part of this
     latter population appears scattered over a wide area. We
    explore several ideas to explain the origin of this relatively young
   component in the inner halo of NGC~253.}
{}


   \maketitle
%

\section{Introduction}

The impressive evidence of stellar streams in the outer regions of
galaxies \citep[e.g.][]{ferguson+02, md+08, md+09, md+10, mouhcine+10,
  chonis+11,miskolczi+11} dramatically shows the importance of merging in their
formation, and at the same time discloses how the study of the stellar
populations in the halos of galaxies can be effective for
understanding the galaxy formation process.  Indeed, although the halo
contains only a small fraction of the mass of the galaxy, the long
dynamical times keep the memory of the assembly history of the galaxy
over a considerable fraction of its life.  It is then of great
interest to probe the extent of the halo, its structure and amount of
substructure, and its stellar population content.

The panoramic view of the M31 - M33 complex \citep{mcconnachie+09} has
revealed the presence of several substructures, scattered
over the whole probed volume, that are remnants of interactions of M31
with its less massive neighbours.  A similar situation likely holds for
the Milky Way, as indicated by the presence of the Sagittarius Dwarf
\citep{ibata+94}, and other structures \citep[e.g.\ ][and references
therein]{yanny+03,ibata+03,belokurov+06,juric+08,grillmair09}.  For
both the Milky Way and the Andromeda galaxy, these substructures seem
to reside above a diffuse halo, which can hardly have originated from
the late disruption of dwarfs, due to the very long dynamical
timescale in such low density regions of galaxies.  The presence of
this underlying halo is particularly puzzling in the frame of the
current picture of galaxy formation in a Cold Dark Matter universe,
and substantial effort is invested to hunt for this elusive component
in nearby galaxies
\citep{rejkuba+09,jablonka+10,tanaka+11,barker+12, cockcroft+13,
  monachesi+13}.

The amount of substructure in the halo traces its buildup through
hierarchical accretion and merger episodes. In the Milky Way,
\citet{bell+08} measured 30-40\% RMS fluctuation in the stellar
distribution of the halo stars with respect to a smooth model, while
\citet{starkenburg+09} estimated a minimum fraction of accreted stars
in the halo of 10\%. Evidence of incompletely mixed sub-populations
was also found in the inner halo of NGC~891, where the RMS fluctuation
in stellar distribution is 14\% \citep{ibata+09}.  An additional
feature which is gaining reliability in the literature is that the
stellar halo may actually consist of two components, one more
internal, elongated, and relatively metal rich, and one external,
almost spherical and metal poor \citep{carollo+07}.

On the theoretical side, sophisticated galaxy formation models in a
$\Lambda$CDM universe have been developed, with specific predictions
about the stellar population properties and surface brightness
profiles which can be compared to observations
\citep{abadi+06,zolotov+09,cooper+10, font+11}.

An analysis of the simulated stellar halos of \citet{cooper+10}, that
were formed via satellite galaxy accretions within the cosmological
N-body simulations of the Aquarius project, showed a systematically
larger RMS in the stellar density distribution with respect to the one
measured for the Milky Way halo by \citet{bell+08} on the Sloan Digital Sky Survey data.
Based on that \citet{helmi+11} estimated that $\sim 10$\% of the Milky
Way halo stars formed \textit{in situ}.  According to \citet{font+11}
simulations, who instead used the GIMIC suite of cosmological
hydrodynamical simulations, the inner halo stars are predominantly formed
\textit{in situ}, while the outer regions of the halo should be mostly
populated with stars originating from accreted satellites.  The
\textit{in situ} component in these simulations comes from a proto-disk
formed at high redshift, which later had its stars dispersed to form
the inner halo via dynamical heating associated with mass accretion
\citep{mccarthy+12}.  As a consequence, these models are characterized
by a change of slope of the surface brightness increase with radius,
which occurs at about 30 kpc for a Milky Way type of galaxy. The
different proportions of \textit{in situ} star formation would also
account for a general metallicity gradient with high metallicity stars
found preferentially in the inner regions.

Studying the stellar distribution in the outskirts of galaxies has
great potential to constrain these models. To this end, the best
suited objects are (massive) spiral galaxies viewed edge on, because
of the favorable geometry to trace the stellar halo.  NGC 253 is one
such galaxy. It is sufficiently near to allow us to map its oldest
stellar population, traced by bright Red Giant Branch (RGB) stars. The
wide field of view of the VISTA telescope offers the opportunity to
study the stellar content in the outskirts of its disk and in its halo
in one shot. Notice that the extended halo is characterized by very
low surface brightness, down to $\mu_{\rm V} \simeq $ 33 mag/square
arcsec or fainter. Reaching these levels (many magnitudes fainter than
the sky) with surface photometry is extremely challenging
\citep{zackrisson+12}, while photometry of individual stars has proved
very successful \citep[e.g.][]{ibata+07}.

In these kinds of studies foreground and background contamination is an important
issue, due to the intrinsically low surface density of the sought after 
stellar halo. In this respect, NGC 253 is favourably located at high
Galactic latitude ($\mathrm{b}=-88^\circ$), such that the foreground 
extinction is very low ($\mathrm{E(B-V)}=0.02$) and we expect the 
foreground sources to be fairly uniformly distributed in the VISTA field 
of view.  This is an advantage
with respect to studying the halos of large Local Group galaxies, where the wide
solid angle is more likely to intercept local structures and
foreground inhomogeneities.

\begin{figure}
\centering
\resizebox{\hsize}{!}{
\includegraphics[angle=0,clip=true]{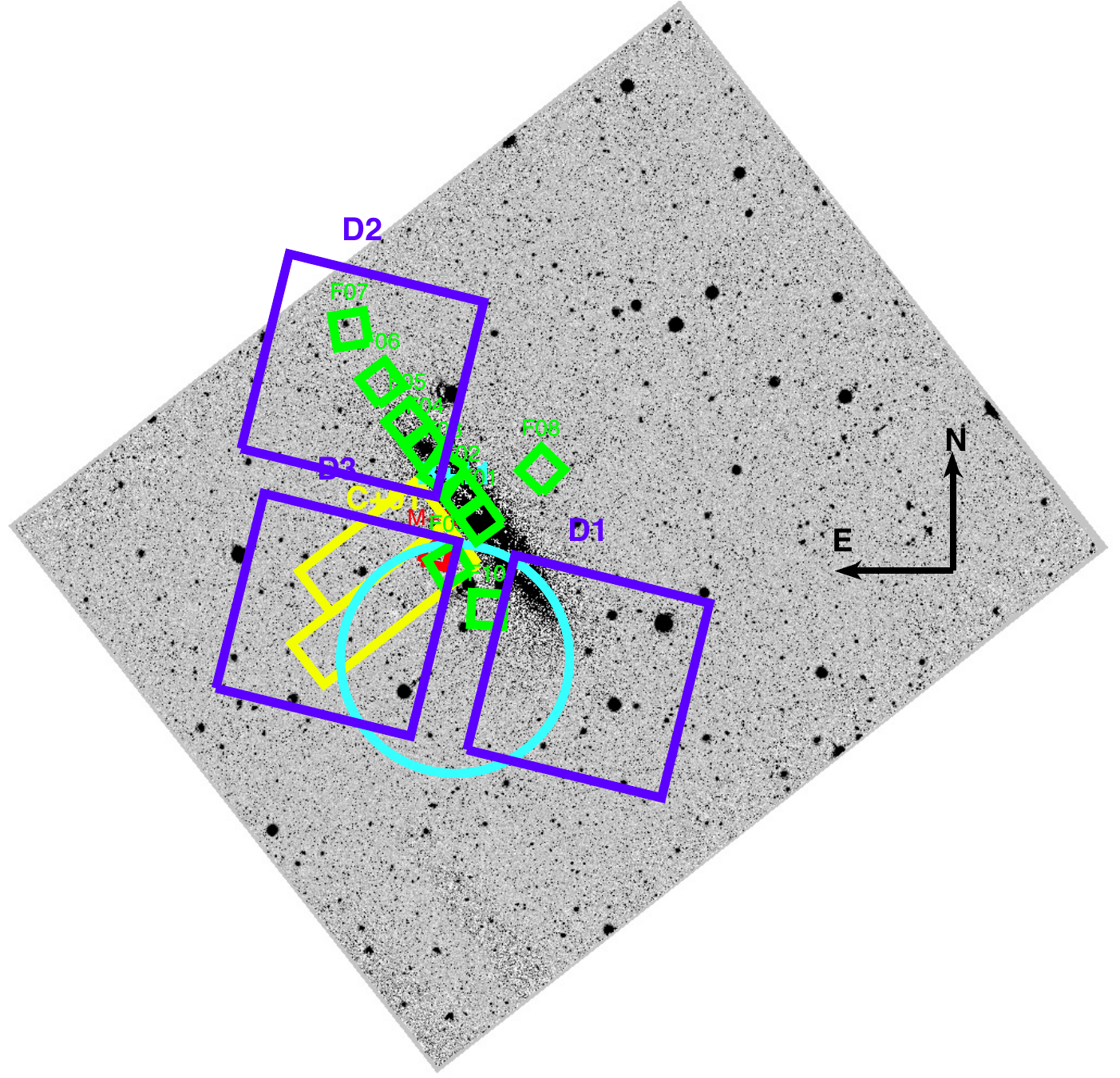}
}
\caption[]{VISTA image of NGC~253 with over plotted the field of view
  of previous resolved stellar population studies from the literature;
  specifically IMACS@Magellan \citep[][cyan circle]{bailin+11}, 3
  WIRCam@CFHT fields \citep[][large blue squares]{davidge10}, NTT
  SUSI2 \citep[][yellow rectangles]{comeron+01}, and many ACS fields
  \citep[GHOST survey,][small green squares]{rs+11}. The HST WFC2
  field from \citet[][small red square]{mouhcine+05I} partially
  overlaps with ACS, SUSI2, and IMACS fields. The wide coverage of the
  VISTA data can be readily appreciated.}
\label{fig:n253_fov}
\end{figure}

NGC~253 is the brightest ($M_{\rm B} \sim -$20) member of the Sculptor
group, which is an extended filament of galaxies with distances ranging from
$\sim$ 2 to 4.5 Mpc \citep{jerjen+98,karachentsev+03}.  NGC~253 is a
luminous infrared source which hosts a recent starburst in its central
part and sustains a galactic wind \citep{rieke+80,rieke+88}. Its inner
regions and X-ray properties were studied by many authors \citep[e.g.\
][and references therein]{fabbiano+84,westmoquette+11}.  The strong
nuclear outflow in NGC~253 is traced by huge lobes of diffuse X-ray,
UV, H$\alpha$, HI, and far-IR dust emission that extend up to$\sim$9
kpc away from the disk
\citep{strickland+02,bauer+08,hoopes+05,boomsma+05,kaneda+09}.

This outflow is driven by widespread star formation activity, which
is not only confined to the nuclear starburst but is also evidenced by
bubble like structures visible on the disk \citep{sofue+94}, as well
as by the young supergiant stars in Colour-Magnitude Diagrams (CMDs)
of resolved stars
\citep{dalcanton+09,davidge10,rs+11}. \citet{comeron+01} find young
stars along the minor axis, up to 15~kpc away from the galaxy plane,
suggesting local star formation in the outflow \citep[see
also][]{comeron+03}.  This strong activity may be surprising given
that there is no close companion, but NGC 253 shows clear signs of
interaction, as kinematically distinct structures in the centre of the
galaxy \citep{prada+98}, extra-planar gas and stars, and pronounced
asymmetries in the stellar distribution \citep{davidge10}. The
evidence suggests that this galaxy has experienced a secondary
event during its formation.

Pencil beam, deep HST observations of NGC 253 have been made by
\citet{mouhcine+05I,mouhcine+05III} with the WFPC2 camera, while the
ANGST \citep{dalcanton+09} and the GHOSTS teams \citep{rs+11} observed
with the ACS. The location of the tip of the RGB in the ACS data
indicates a distance modulus of 27.7, slightly larger but fully
consistent with the distance determined by \citet{mouhcine+05I} from
WFPC2 data ($27.59 \pm 0.06 \pm 0.16 {(\it syst)}$).  Some of these
HST fields are located in the periphery of the disk, but are close
to it, while the outer regions of the galaxy were more extensively
probed with wide field imaging from the ground. \citet{malin+97}
traced an extended low surface brightness envelope of about 28
mag~arcsec$^{-2}$ out to $25'$ (corresponding to $\simeq  25$ kpc) distance around NGC~253,
noticing the asymmetrical nature of this stellar halo, and concluding
that it might have been distorted by the infall of a companion
galaxy. Deeper imaging that resolved individual stars in the halo were
published by \citet{davidge10}, and \citet{bailin+11}.  The former
study is not deep enough to sample the RGB stars, so that most of the
information concerns the disk and the extra-planar young and
intermediate-age component, as traced by young red supergiants and
asymptotic giant branch (AGB) stars. This study reveals that the disk
of NGC~253 is disturbed. \citet{bailin+11} are instead able
to trace the RGB population in a wide region which comprises the inner
halo for which they find a rather flattened geometry. Both studies
find a stellar over density in the south, outer disk region,
confirming early finding by \citet{beck+82}.
  
In this paper we present deep $Z$ and $J$ band photometry of the outer
disk and halo of NGC~253 obtained with the VISTA telescope in the
framework of the Science Verification Programme.  In
Fig.~\ref{fig:n253_fov} we provide an overview of the several
pointings from the literature, which were used to investigate resolved stellar
populations in NGC~253, superimposed on the VISTA image. In
comparison to these previous works, our data offer the opportunity to
study the whole region around NGC~253, yielding a wide and continuous
view of the halo out to a distance of $\sim$ 50~kpc.  In
Section~\ref{sec:data} we describe the data, and in Section~\ref{sec:cmd}
present the Colour-Magnitude Diagram (CMD) constructed with the
stellar source catalogue. In Section~\ref{sec:structure} we discuss the stellar
disk and the halo components of NGC~253 plus the perturbation to the
disk, i.e. the southern shelf, and report about the detection for the
first time of a substructure in the halo, at about 30 kpc from the main
plane of the disk. In Section~\ref{sec:SFH} we discuss the star
formation history in the outer disk and halo of NGC~253, and the
spatial distribution of both AGB and RGB population in
these two components. Our results are summarized and discussed 
in Section~\ref{sec:discussion}, and
conclusions are drawn in Section~\ref{sec:conclusions}. Throughout this paper
we adopt a distance modulus to NGC~253 of $27.7$ mag, which is equivalent
to a distance of $3.47$ Mpc, yielding an image scale of $ 16.8$
pc per arcsecond.

\section{The data}
\label{sec:data}

The data presented in this paper were collected as part of the VISTA
Science Verification (SV) programme \citep{arnaboldi+10} which was
designed to test the performance of the integrated telescope-instrument
system with two projects: (i) one aimed at deriving very
deep images in the smallest contiguous area that uses all VISTA's
large field of view (a \textit{tile} in VISTA
nomenclature) whose results are the subject of this paper; (ii) the
other covering a 30 deg$^2$ area around Orion Belt stars with a
mosaic of 20 tiles having shallower exposures
\citep{petr-gotzens+11}. The two projects required different observing
strategies, therefore providing tests of observing templates and
operation tools for the public surveys which started on the VISTA
telescope immediately after the SV \citep{arnaboldi+12}.

\subsection{Observations} 

\begin{figure}
\centering
\resizebox{\hsize}{!}{\includegraphics[angle=0,clip=true]{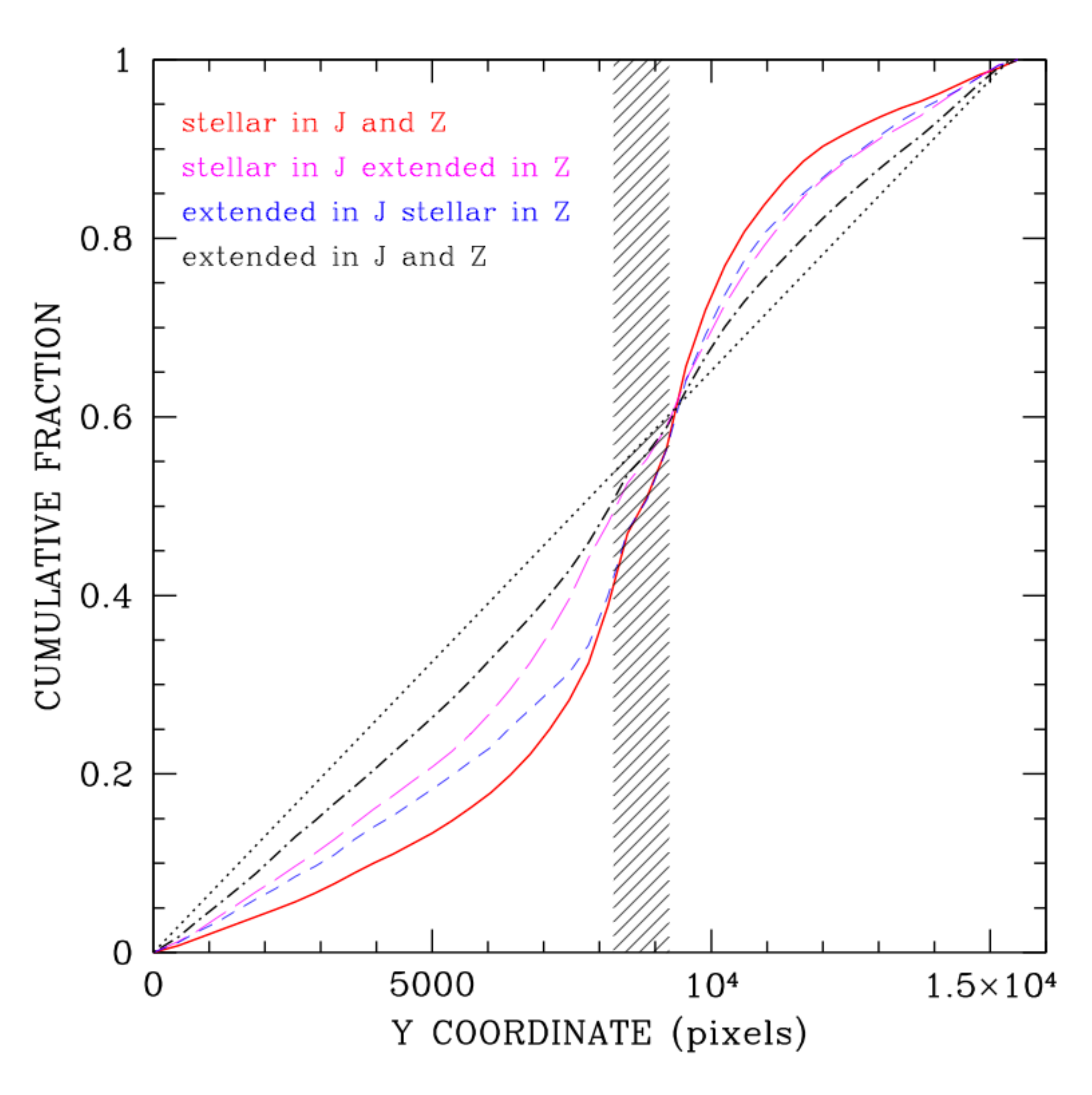}}
\caption[]{Cumulative distribution along the vertical coordinate of
  the VISTA tile for sources with different classifications on the
  VDFS catalogue: stellar in both $J$ and $Z$ (red solid line),
  extended in both $J$ and $Z$ (black dot-dashed line), stellar
  in $J$ and extended in $Z$ (magenta long-dashed line), stellar in $Z$
  and extended in $J$ (blue short-dashed line). A flat spatial distribution corresponds to
  a line at 45 degrees (dotted line). The shaded region indicates the
  location of the disk of NGC 253, where the surface brightness is too
  high to measure individual sources and number counts are heavily affected by
  crowding effects.}
\label{fig:spatial}
\end{figure}

During the SV period (from October 16 to November 2, 2009), NGC 253
was observed in the first part of the night, until the second SV
target, Orion, became observable.

VISTA is a 4m alt-azimuth telescope with a single instrument: VIRCAM
\citep{emerson+06, dalton+06}. The camera has a 1.65 degree diameter 
field of view which is sparsely populated with 16 $2\mathrm{k} \times
2\mathrm{k}$ Raytheon VIRGO detectors with pixels of mean size $\sim 0\farcs34$.
Each individual exposure images a (disconnected) total area of 0.59 deg$^2$.
To sample a contiguous field of $\sim 1.5 \times 1$ deg$^2$
at least 6 individual exposures (so called pawprints) are
necessary with large offsets. Since in this project we aimed to get
deep photometry, the single tile centred on RA=00:46:59.86,
DEC=-25:16:31.8 was imaged many times over the two
week long observing run. Each observing sequence typically consisted
of several (2-5) jittered exposures at one pawprint position, before
making a larger offset to the next pawprint position to cover the
gap. At this second position the same number of jittered
exposures were taken and repeated until six pawprints were accumulated to
fill the tile. These sequences were then repeated with different
filters.

In spite of the large extent of the target on the sky, (NGC~253 is one
of the largest southern galaxies), the observing strategy did not
require extra offset empty sky fields due to the very large size of each 
VIRCAM detector, which covers an area of  $11\farcm 6 \times
11\farcm6$ on sky. The galaxy major axis was oriented at
PA$=51.95^\circ$, aligned with the shorter side of the tile, such that
the galaxy covered detectors 10 and 11 in pawprints 2, 4, and 6, while
it was kept in the gap between the detectors in the other three
pawprints. Hence the jittered exposures of the odd pawprints could be
used to create sky images for the data processing, while at the same time 
being used in the final deep stack.

We acquired deep images of NGC 253 in three filters: $Z$, $J$ and in
NB118. The latter is a narrow band filter centred on the redshifted
wavelength of H$\alpha$ emitters at redshift $z \sim 0.84$ and 
Ly$\alpha$ emitters at $z \sim 8.8$ 
\citep{milvang-jensen+13}.  While the data taken with that filter were
focused on the high redshift universe behind NGC 253, with the deep
$J$ and $Z$-band images we explore the stellar population content and
the structure of the halo of NGC 253. The J-band images consisted of 5
exposures, each with a detector integration time (DIT) of 45 sec. The
Z-band images had $3\times 60$~sec exposure, while several different
DITs were used for NB118 images: 270, 450 and 860 sec. The final 
deep J-band stack was made combining 20 tile sequences of 6 pawprints 
with a varying number of jitters that add up to 22.1h total accumulated exposure
time. The total exposure time of all Z-band sequences was 9.6h, and 
5.8h were accumulated with the NB118 filter. On target exposure time is 1/3 of the 
total accumulated exposures due to large offsets necessary to fill the gaps between
detectors. 
These integration times allowed us to reach J=23.5 and Z=24.5 (on the Vega magnitude system) with a S/N of 3.

In addition, we acquired shallower exposures in all the VISTA
broad-band filters (Z, Y, J, H and Ks) with the aim of studying the
central part of the disk of NGC~253 and its structure.  In this paper
we analyze the deep Z and J-band images, while the shallow broad-band
imaging data are described in a subsequent paper (Iodice et al., in
prep).  

During the observing run the seeing and sky transparency
varied, but the conditions were generally very good, with only two
nights having thin clouds (but excellent seeing) and two nights being
lost due to strong wind.  The final deep
stacked image has an average seeing of $\sim 0\farcs9$ in the $J$-band
and $\sim 1\farcs 2$ in the $Z$-band.

Figure~\ref{fig:n253_fov} shows the final stacked J-band image, mapped
with $\simeq 13000 \times 16000$ pixels.  At the distance of NGC~253
the image samples an area of about 74~kpc$ \times$ 91~kpc around 
NGC~253. 
\begin{figure*}
\centering
\resizebox{\hsize}{!}{
\includegraphics[angle=0,clip=true]{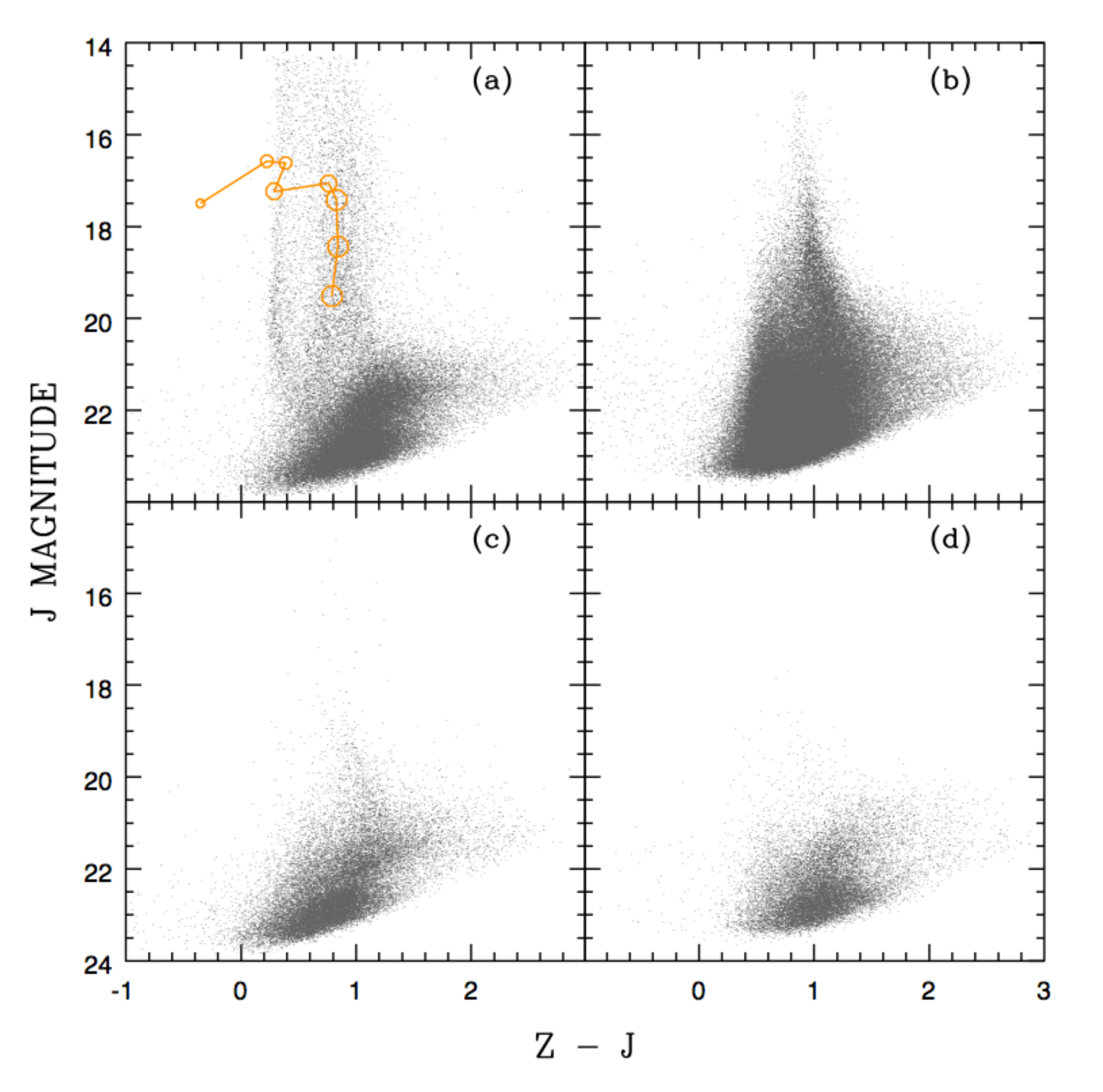}
}
\caption[]{Observed CMDs for objects with different classifications
  from the reduction package: stellar sources on both filters (panel
  (a), 55025 points); extended sources in both filters (panel (b),
  151924 points); stellar on the $J$ and extended on the $Z$ tile
  (panel (c), 27687 points); and extended on the $J$ and stellar on
  the $Z$ image (panel (d), 18932 points). In panel (a) the open (orange)
  circles show the evolutionary path of a simple stellar population
  (SSP) model with metallicity $Z$=0.008 as the population ages. The
  size of the circles increases with the age of the model plotted at
  (4, 10, 30, 100 and 300) Myr and at (1, 3 and 10) Gyr. The model was
  obtained using the CMD on-line tool by L. Girardi at
  stev.oapd.inaf.it/cmd, and refers to an SSP born with 10$^5$\msun\
  of stars distributed between 0.1 and 100 \msun\ with a Salpeter IMF
  flattened below 0.5 \msun.  }
\label{fig:cmd_class}
\end{figure*}

\subsection{Data Reduction and Photometry}
\label{sec:reduction}

Data reduction was carried out with the VISTA Data Flow System 
\citep[VDFS,][]{lewis+10} at the Cambridge Astronomy Survey Unit (CASU). The 
reduction procedure consists of the following standard steps: 
dark correction which removes the dark current and also corrects other 
additive electronic effects;
linearity correction to account for the non-linear overall response of
the VISTA detectors;
flatfield correction using stacked twilight flats which is also used to
gain-normalise all the detectors to a common internal system;
and sky background correction to remove the large-scale spatial background
variation from atmosphere variations.
In addition, the VISTA detector IRACE controllers imprint a low-level 
horizontal stripe pattern on the images. While the position and amplitude 
of the stripes varies from one exposure to the next, it is the same along the 
row across the four detectors read out through each IRACE controller.
Therefore, even though the extended and bright galaxy was occupying the whole 
of detector number 10 and a large fraction of detector 11, the other two 
detectors could be used to remove the striping pattern from the galaxy images. 
The pipeline then combines the individual jittered exposures for each OB to 
make deeper individual pawprint stacks, and the six stacked pawprints are 
combined to produce a tile image (and catalogue) for each OB. 

The final deep stacked tile image and catalogue used for the science 
analysis was created by first combining all suitable sets of 6 pawprint 
stacks to make deeper pawprint stacks.  Prior to making the final tile image
any background variation in the pawprint component images was removed using 
the so-called nebulosity filter developed by CASU\footnote{see
http://casu.ast.cam.ac.uk/publications/nebulosity-filter}.  This filter 
differentially removes all smoothly varying background on a specified scale, 
in this case set to $\approx$30 arcsec.  In addition to generic background 
variations the filter also drastically reduces the impact of large-scale 
reflection halos around bright stars and also removes most of the unresolved 
disk/halo light from NGC~253.

These "nebulised" pawprint images were then combined to make the final deep 
tile image. A few individual OB exposures were affected by electronic noise 
appearing occasionally on channel 14 of detector 6. However, the total area 
affected by that extra noise was less than 0.4\% of the whole tile, and as
most images did not have this extra noise present this was effectively
removed during the data reduction.  In contrast the upper half of detector 16 
(which covers the southern corner of the tile in Fig. 1) is unstable, 
particularly in the bluer bands, and was masked off in the confidence maps
prior to making the final tile images. Although the whole tile image is
available, objects falling on this part of detector 16 have shorter effective
integration times, by a factor 2, compared to the rest of the tile resulting
in shallower photometry in the affected regions.  Finally, prior to making
the deep Z- and J-band catalogues the central region of NGC253 was masked out. 
No other masks, e.g.\ for bright foreground stars, were used.

The cataloguing and photometric measurements were done with the VDFS
pipeline using the standard CASU imcore package. This package measures
fluxes in a range of apertures and enables shape characterization
for all objects.  The deblending option was used to enable better
measures of stellar sources in the crowded outer disk regions of the galaxy. 
The shape parameters and series of apertures fluxes were used to generate
morphological classification information and to compute stellar aperture
corrections.  Photometric (and astrometric) calibration was based on 
unsaturated 2MASS stars converted to the VISTA (Vega-like) system.

The photometric errors as a function of magnitude are well fitted with
an exponential function:
\begin{eqnarray}
\sigma_J = 5.58\times 10^{-10} \times \mathrm{e}^{0.803 \times J + 1.17} \\
\sigma_Z = 1.87\times 10^{-9} \times \mathrm{e}^{0.774 \times Z  -0.026}  \  .  
\end{eqnarray}

\begin{figure}
\centering
\resizebox{\hsize}{!}{\includegraphics[angle=0,clip=true]{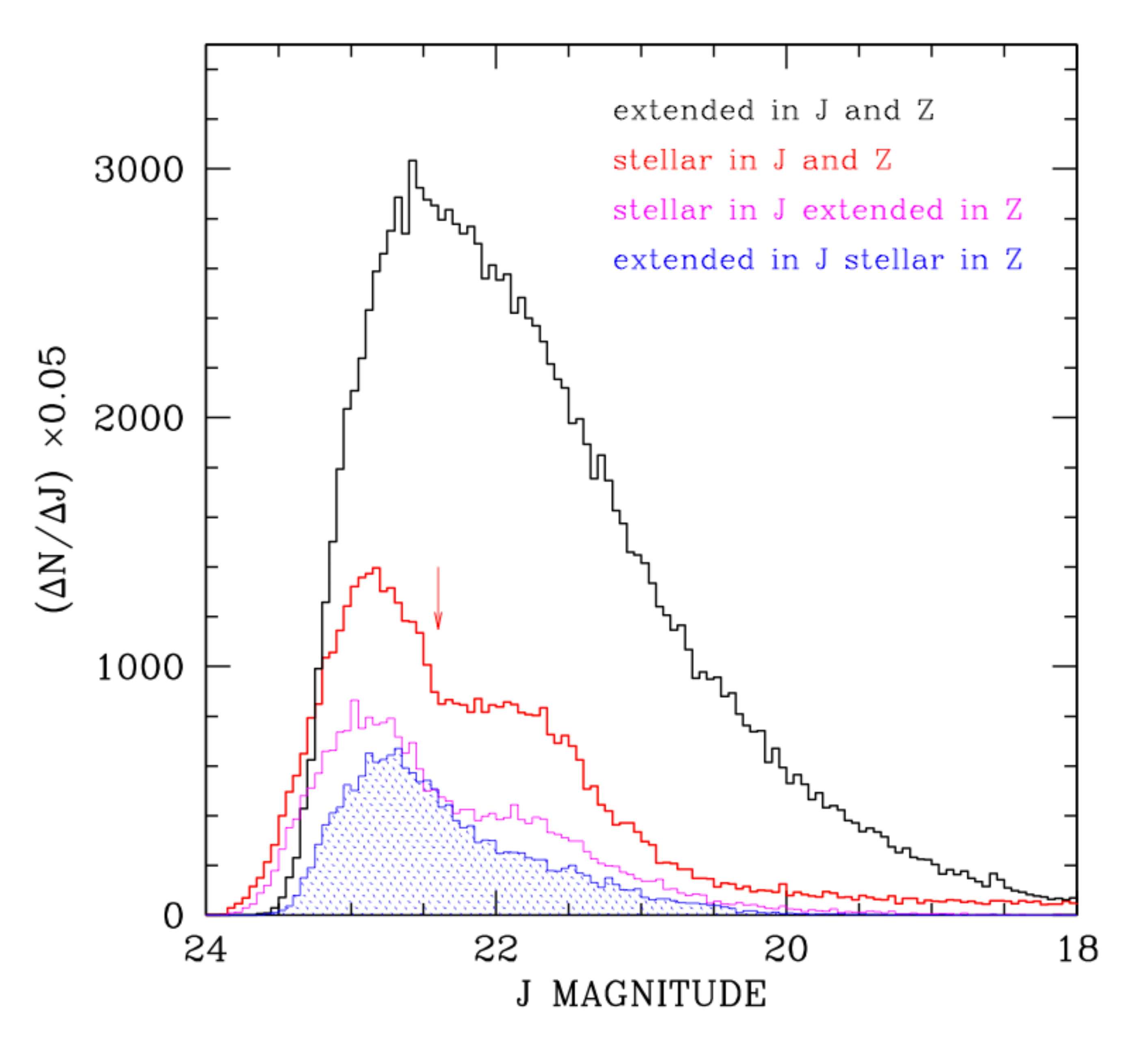}}
\caption[]{Luminosity functions of the different kinds of sources
  whose CMDs are shown in Fig. \ref{fig:cmd_class}. The red thick line shows the 
  distribution of sources classified as stellar in both filters corresponding to 
  panel (a) and the black thick line is used for the sources classified as extended
  in both filters, corresponding to the panel (b). The 
  thin magenta line and the shaded blue histograms refer to the sources
  shown in panels (c) and (d) of Fig. \ref{fig:cmd_class}. Labels in the 
  figure follow the order of the histograms from top to bottom.
  The arrow indicates the magnitude at which we
  expect the Tip of the RGB for a stellar population with metallicity
  $Z=0.008$, and a distance modulus of $27.7$.}
\label{fig:lf_class}
\end{figure}

The catalogue from the VDFS pipeline includes more than $400,000$
sources matched on the $J$ and on the $Z$ tiles, but $\sim 30$\% 
of them are either spurious sources, or their magnitude and/or shape are
very uncertain.  Most of these objects are classified as noise in
either band, and a few of them are instead saturated or contain a bad
pixel. We do not consider these detections in the following
analysis. In addition, for $\sim 10$\% of the remaining objects,
the positional coincidence of the source on the two tiles is
particularly poor. We also discard these cases, since we aim at the
construction of a  catalogue of NGC~253 bona fide star members, and the
poor quality of the positional match may indicate the presence of an
extended object or a blend of two or more stars. After this cleaning,
the catalogue includes about $253,500$ sources classified as
\textit{stellar} or \textit{extended} in either the $J$ or $ Z$ tile,
or both.  Inspection of the spatial distribution of the various
categories of sources shows that all of them are concentrated toward
the disk of NGC~253. This is due to crowding, since the de-blending of
stellar sources becomes progressively more difficult as the surface
brightness increases. This effect enhances the number of sources
classified as \textit{extended} as the galaxy disk is
approached. However, the spatial distribution of these four kinds of
detections show quantitative differences, as illustrated in
Fig.~\ref{fig:spatial}. Sources classified as extended on both tiles
deviate the least from a flat distribution, while sources classified
as stellar on both tiles deviate the most, with a strong excess in the
shaded region, where the disk is located. Sources classified
differently on the two tiles exhibit a behavior which is
intermediate between the flat distribution and the population
concentrated toward the disk of NGC~253.

In Fig.~\ref{fig:cmd_class} we plot the CMD
of these four kinds of sources. In panel (a) we plot the CMD for
sources that are stellar in both J and Z; here the stellar population
of NGC~253 is clearly visible at magnitudes fainter than $J \sim$ 20,
with the vertical plumes of foreground stars superimposed. In panel
(b) we plot the CMD of sources extended in J and Z. This CMD is very
different from that in panel (a), suggesting that the majority of the
sources classified as extended in both filters are background
galaxies.  The CMDs shown in panels (c) for the objects that are
stellar in J and extended in Z, and in (d) for objects extended in J
and stellar in Z contain mostly faint sources, whose light
distribution has lower S/N. In Fig.~\ref{fig:lf_class} we show the
luminosity functions (LF) of the four kinds of sources. It appears
that objects with the same morphological classification on the $J$
tile have a similar magnitude distribution. In particular, the LFs of
the sources classified as stellar on the $J$ band image (plotted in red and magenta) present a discontinuity at $J \simeq 22.5$ where the Tip
of the RGB is expected if a distance modulus of 27.7 is applied to the 
Padova isochrones.  We then conclude that most of the
sources in panel (c) of Fig.~\ref{fig:cmd_class} are likely stars, and
they appear extended on the $Z$ image because of the worse
seeing. Differently, most of the sources in panel (d) are likely to be
compact background galaxies.

Because NGC~253 is undergoing a starburst, we consider the presence of
(young) globular clusters in our area.  In panel (a) of
Fig.\ref{fig:cmd_class} we show the evolutionary track appropriate for
a globular cluster of $10^5$ \msun\ with metallicity similar to that
of the LMC; for a solar metallicity, such path would be shifted to
redder colours by $\sim 0.15$ mag. While it is possible that some
clusters are present in the CMD of the stellar sources, confused among
the foreground stars, the CMDs of extended sources in both
bands are very unlikely to contain young and blue (globular) star clusters.  
Indeed, we do expect globular clusters to appear as 
stellar-like sources in our image, because the median effective radius
of Milky Way Globular Clusters is $3.2$ pc \citep{harris96}, which
corresponds to $\sim0\farcs2$ at the distance of NGC 253,
well within the seeing disc.

In what follows, we consider as bona fide stars the sources classified
as stellar in both filters, and whose CMD is plotted on panel (a);
this is our catalogue A. However, a fair fraction of the sources
classified as stars in $J$ and extended in $Z$ could be stellar
members of NGC 253 also. We take these sources into account when
discussing the halo of the galaxy, its extension and shape. We refer
to the cumulative catalogue as catalogue B.

\begin{figure}
\centering
\resizebox{\hsize}{!}{
\includegraphics[angle=0,clip=true]{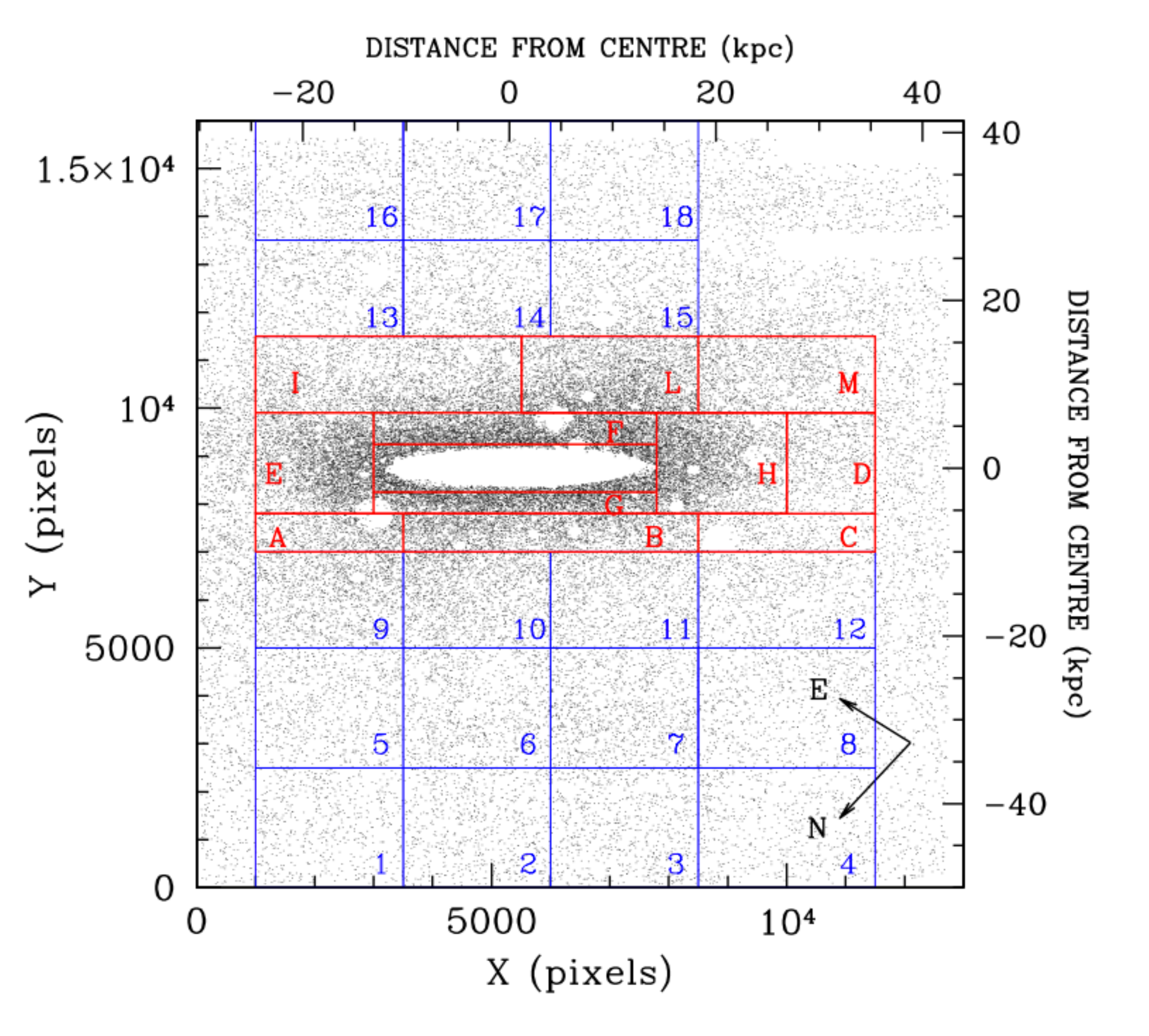}
}
\caption[]{Spatial distribution of catalogue A stars on the VISTA tile
  ($J$ band) oriented along the detector coordinates. The upper and right axis are labelled with physical
  coordinates in kpc centred on the galaxy centre. Holes are clearly visible in the spatial distribution.
  These are due to the masking of bright foreground stars. The inner
  part of the disk has also been masked, because source crowding
  prevents accurate photometry on our images. In the upper right
  corner, two white rectangles show other regions with defects, where
  the photometry could not be performed. The cumulative area of the
  masked regions is a small fraction of the total surveyed area. Solid
  lines show the rectangular sub-regions into which we divide the tile to
  draw the spatially resolved CMDs shown on Fig.~\ref{fig:stamps_halo}
  and Fig.~\ref{fig:stamps_disc}. The sub-regions are drawn so as to
  sample a statistically significant population of stars, and are identified with 
  progressive numbers and letters, for the halo and the
  disk regions respectively. }
\label{fig:map}
\end{figure}

\section{The Colour-Magnitude diagram of stellar sources in NGC
  253}\label{sec:cmd}
 
\begin{figure}
\centering
\resizebox{\hsize}{!}{
\includegraphics[angle=0,clip=true]{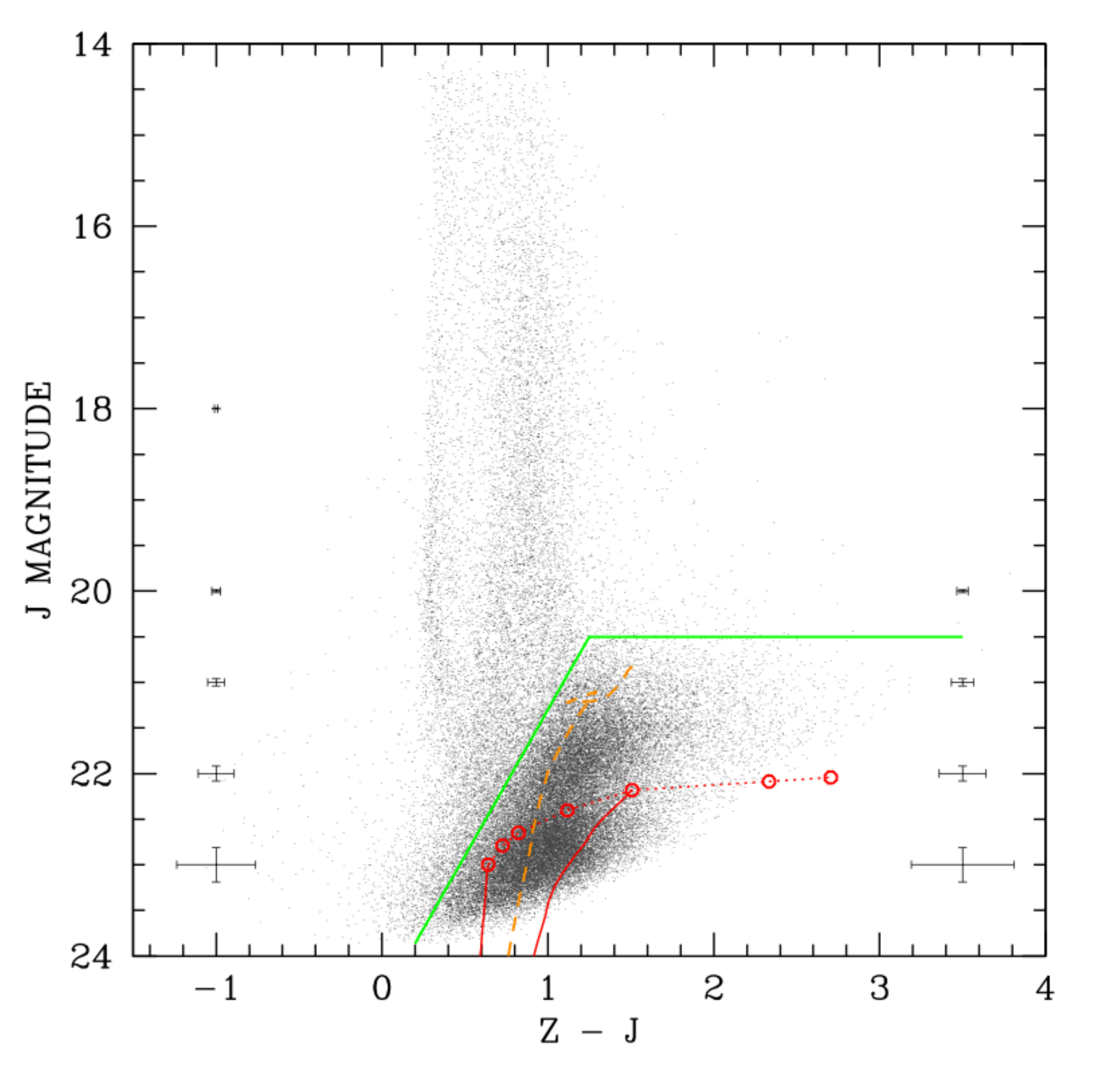}
}
\caption[]{CMD of catalogue A stars 
  with theoretical loci superimposed. The models are
  computed with the CMD tool at stev.oapd.inaf.it/cmd and shifted by a
  distance modulus of 27.7. Open circles, connected with a dotted line, show the location of the RGB
  Tip of 10 Gyr old isochrones with metallicity Z=(0.0001, 0.0004,
  0.001, 0.004, 0.008, 0.019 and 0.03), falling, respectively, at
  $J$=(23, 22.79, 22.65, 22.4, 22.18, 22.09 and 22.04).  Solid thick
red lines show the RGB portion of 10 Gyr old isochrones with Z=0.0001
  and Z=0.008; the dashed orange line shows a 1 Gyr old isochrone with
  Z=0.008. The green solid line shows the boundary below which most
  stellar sources are likely members of NGC 253 (see text). 1$\sigma$ error bars are shown
  for two values of the star's color: those  on the left refer to $J-Z=0.5$, while those on the right to
  $J-Z=1.2$ mag.}
\label{fig:cmd}
\end{figure}

Fig.~\ref{fig:map} shows the spatial distribution of catalogue A
stars. An extended stellar disk is easily identified at $1000 \lesssim
X \lesssim 11000$, $7000 \lesssim Y \lesssim 11000$, behind a
sheet of foreground stars  and unresolved background galaxies. 
The disk of the galaxy falls in the upper left section of the
tile, with the centre located at $X \simeq 5300, Y\simeq
8750$. Therefore we probe the galaxy disk up to a distance of more
than 40~kpc along the major axis (South-West direction), and
the halo up to a distance of $\simeq$50~kpc, along the minor axis 
(North-West direction).
The inner part of the disk has been masked and shows up as
white region on the figure. Few very bright foreground Milky Way
stars are also visible as white "holes". 
Two rectangular white stripes in the upper right 
corner are due to the lower sensitivity of detector 16 (Sec.~\ref{sec:reduction}), 
and this region is not considered in the further analysis.
The distribution of the stars on the
disk appears asymmetrical, with an excess of sources on the
upper/right side of the disk.  This is the well known \textit{southern
  shelf} already recognized in the literature
\citep{beck+82,davidge10,bailin+11}. Some other distortions of the
disk may be recognized (e.g. on the left side of the disk along the
North direction), but it is difficult to assess their reliability in this figure due to
the foreground contamination.

The CMD of the bona fide stars  (catalogue A) is shown in
Fig.~\ref{fig:cmd}, superimposed with a few theoretical loci from the
Padova database, shifted to a distance modulus of 27.7. The dotted
line connects the theoretical RGB Tip models of 10 Gyr old isochrones
with metallicity from $\sim$ 1/200 $Z_\odot$ to $\sim 1.5 Z_\odot$.
The stellar population of NGC~253 stands out from the foreground
contamination as a strong enhancement of faint and red sources. The
comparison with the evolutionary models reveals that our CMD contains
a bright RGB component plus an AGB component extending beyond the RGB
Tip (i.e. $J \lesssim$ 22.5). 
The colours of the RGB stars cover a wide range due to a combination of 
a metallicity spread and photometric errors. Using WFPC2 $V$ and $I$ data 
of a halo field of this galaxy \citet{mouhcine+05III} measured a metallicity 
distribution ranging from [Fe/H] $\sim -2.3$ to [Fe/H] $\sim -0.2$. 
The overplotted 10 Gyr isochrones (Fig.~\ref{fig:cmd} ) show
consistency between our data and this metallicity range, taking into
account the photometric errors. 
The luminosity extension of the AGB component indicates the presence of 
intermediate age ($\sim$ 1 Gyr old) stars.

The CMD in Fig. \ref{fig:cmd} shows where the stellar members
of NGC 253 are located. This is the region below the solid green
line. The rest of the CMD is dominated by the foreground component,
and some contamination will also be present in the selected region. In
the next section we assess the contribution from the Milky Way (MW)
stars.

\subsection{Foreground contamination: the TRILEGAL model}\label{trilegal}

\begin{figure}
\centering
\resizebox{\hsize}{!}{
\includegraphics[angle=0,clip=true]{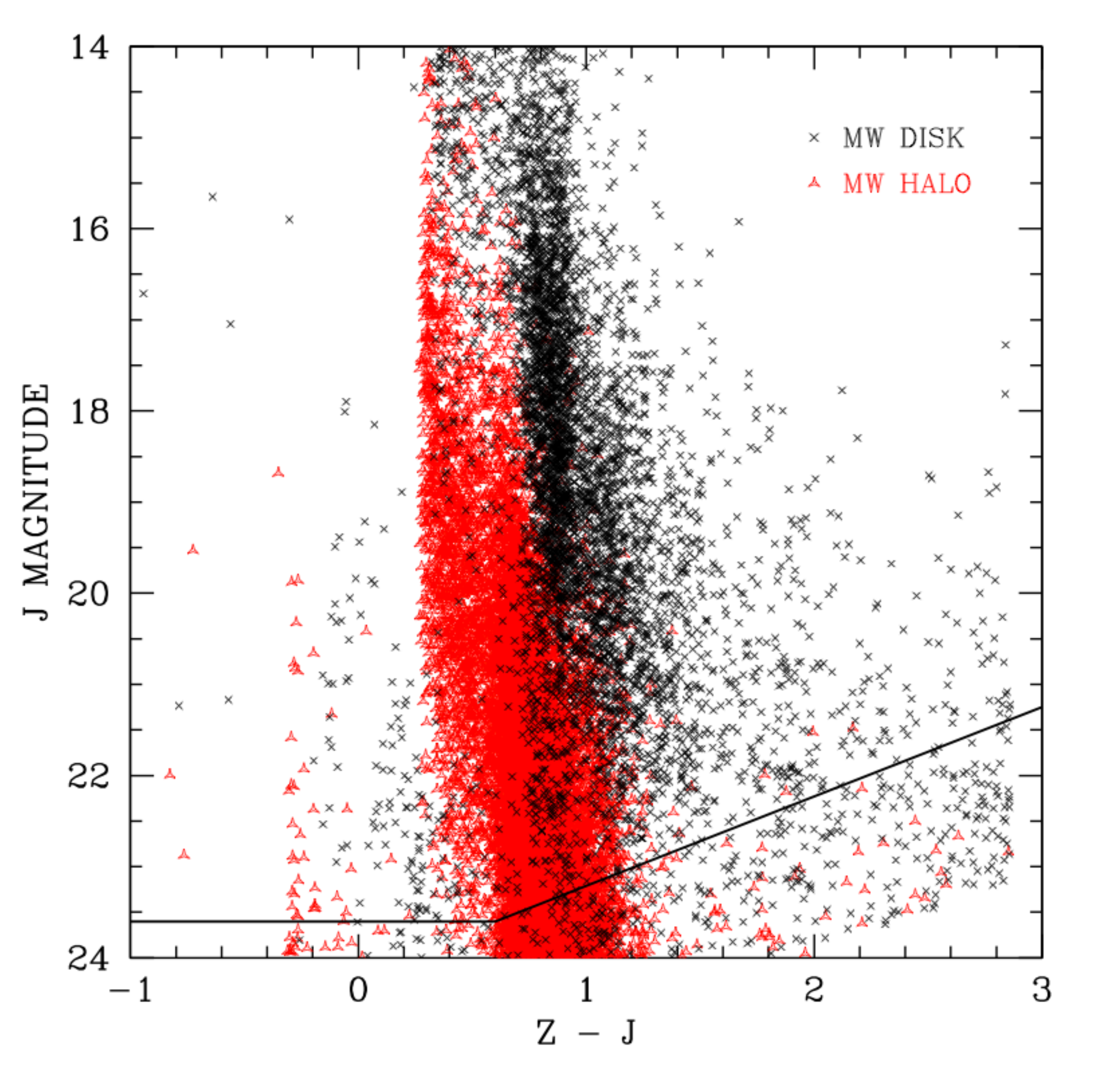}
}
\caption[]{CMD of the expected foreground stellar population as derived 
  from the TRILEGAL simulation. The colour and point type encode the Milky Way
  component, as labelled.  The solid line mimics the faint limit of the observed CMD of
  NGC~253.  The simulation has been calculated for a 1.46 deg$^2$ FoV
  centred at the Galactic coordinates: $l=97.37$, $b=-87.96$,
  adopting a Chabrier IMF, and keeping all other parameters at their
  default values.}
\label{fig:fore_cmd}
\end{figure}

\begin{figure}
\centering
\resizebox{\hsize}{!}{
\includegraphics[angle=0,clip=true]{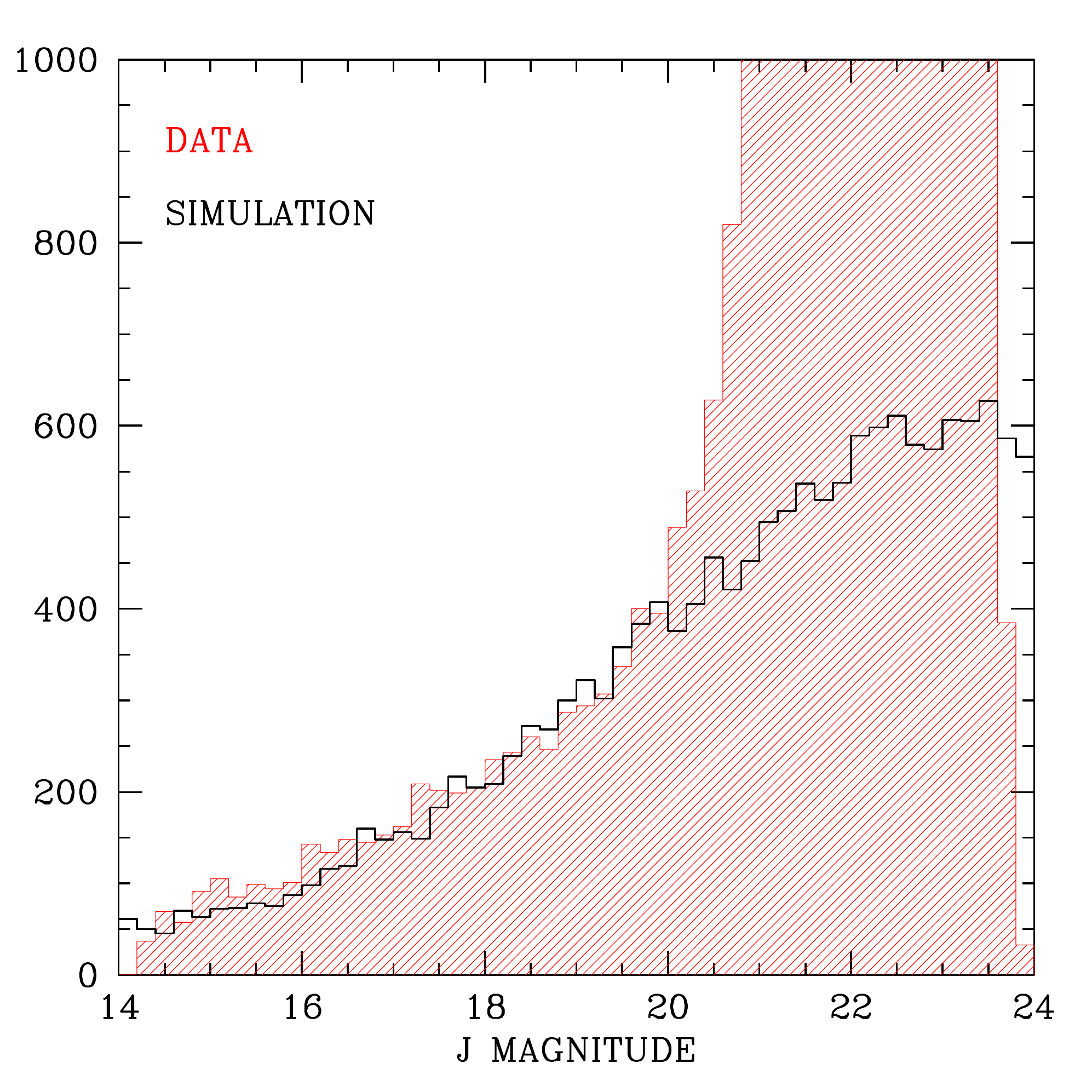}
}
\caption[]{Luminosity function of the expected foreground stellar
  population shown in Fig. \ref{fig:fore_cmd} (black line) from the
  TRILEGAL simulation compared with the data (red shaded histogram). No
  scaling factor has been applied to bring the model and the data into
  agreement.}
\label{fig:fore_lf}
\end{figure}

\begin{figure}
\centering
\resizebox{\hsize}{!}{
\includegraphics[angle=0,clip=true]{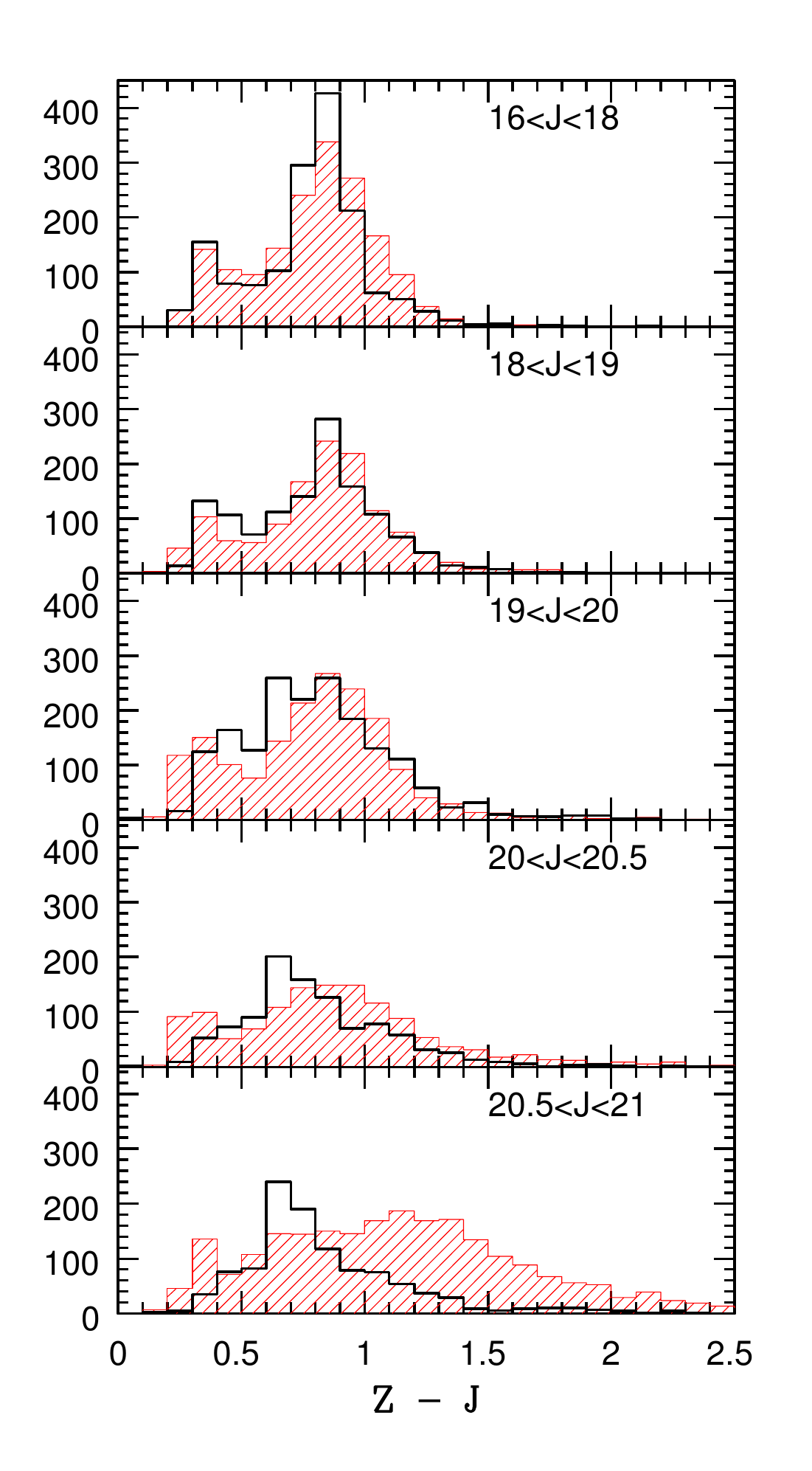}
}
\caption[]{Colour functions in the indicated magnitude
  ranges. The red shaded histograms are computed from the data in catalogue A;
  the solid black lines show the expected foreground stellar
  population as computed using TRILEGAL, see Fig.~\ref{fig:fore_cmd}.}
\label{fig:fore_cf}
\end{figure}

In spite of the high Galactic latitude of NGC~253, the shear size of our
field and its photometric depth provide a
large population of MW foreground stars. This component may be
estimated from the stellar counts at the edges of the tiles, but since
we do not know how extended the halo of NGC 253 is, and measuring its
radius is one goal of the current project, we first evaluate the
foreground contribution in an independent way.
Fig. \ref{fig:fore_cmd} shows the CMD of a simulated population of
foreground stars as obtained using the TRILEGAL tool version
1.5\footnote{stev.oapd.inaf.it/cgi-bin/trilegal} \citep{girardi+05}
for a field with an area of 1.46 deg$^2$ centred on the Galactic
coordinates of NGC~253.  The foreground population consists of a disk
plus a halo component which, according to the simulation, contribute
respectively 36$\%$ and 64$\%$ to the total foreground counts within
the CMD region considered here (i.e. $-1 \leq (Z-J) \leq 3; 14 \leq J
\leq 24)$. The great majority of model stars in
Fig. \ref{fig:fore_cmd} are Main Sequence objects with mass up to
$\sim 1$ \msun, except for the bluest points ($Z-J \la 0.5$) which are
white dwarfs; the red tail of objects at $Z-J \geq 1.2$ are disk red
dwarfs with masses between 0.07 and 0.15 \msun.
Fig. \ref{fig:fore_lf} shows the luminosity function of the simulated
foreground population. The agreement with the observed counts at $J
\lesssim 20$ is remarkable, given that this simulation is not a fit to
our data, and the TRILEGAL default parameters were calibrated on
different data sets. At fainter magnitudes the observed star counts
exceed the simulated ones. 
While there is no guarantee that
 the model is correct at these faint magnitudes, which are poorly
 constrained from observations,  the strong excess
 in our observed luminosity function sets in at a 
magnitude consistent with the  expectation for bright stars at a 
 distance of NGC~253.

\begin{figure}
\begin{center}
\resizebox{\hsize}{!}{\includegraphics[angle=0,clip=true]{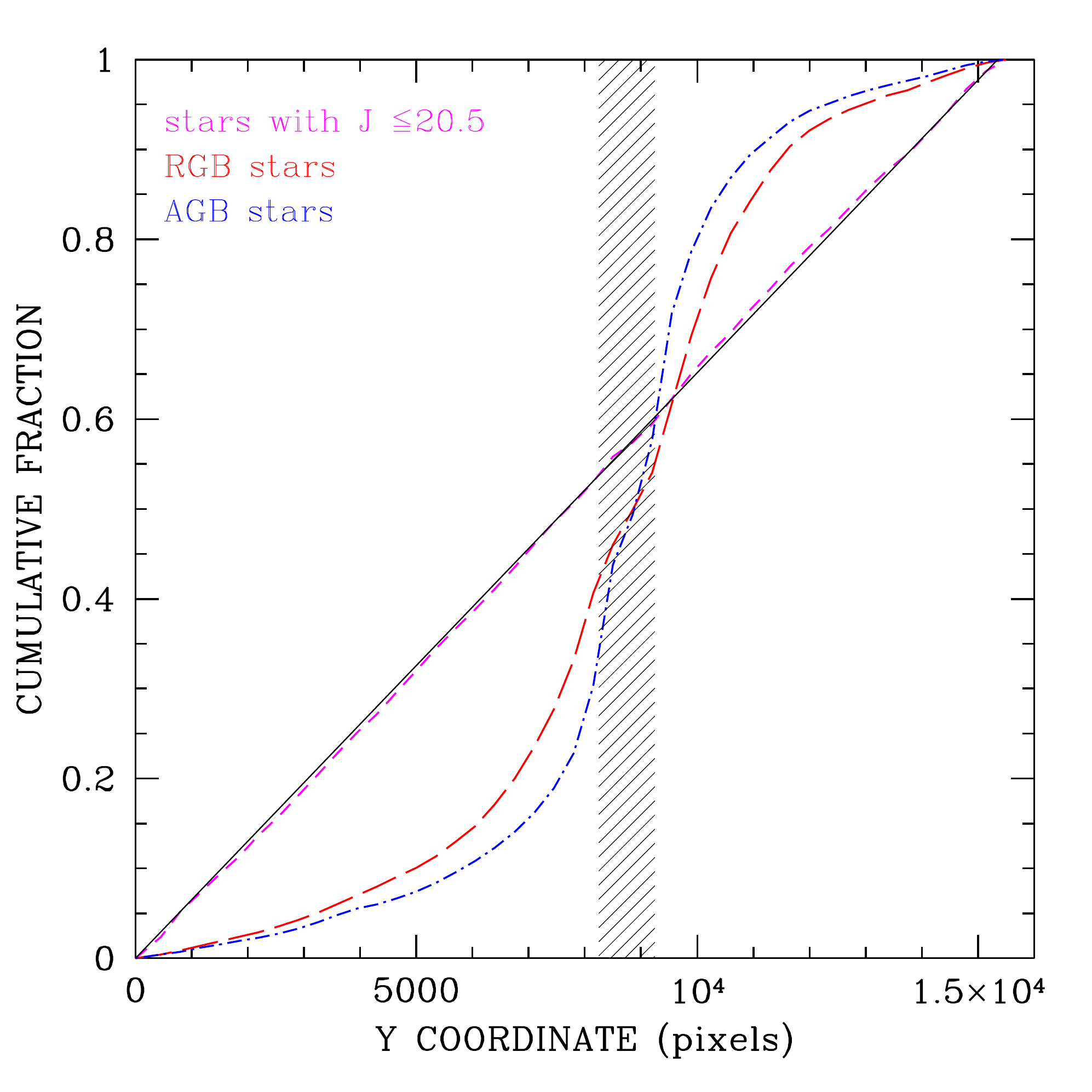}}
\caption{Cumulative distributions along the vertical (Y) coordinate of the
  tile for stellar sources with different locations on the CMD. 
  The short dashed (magenta) line refers to
  stars brighter than J = 20.5;  RGB
  (red, long dashed) and AGB (blue, dot-dashed) stars are located below
    the solid green line in Fig. \ref{fig:cmd},  respectively below and above the dotted red line 
    connecting the RGB tips from evolutionary models. }
\label{fig:spatial_stars}
\end{center}
\end{figure}

To constrain the brightest magnitude at which the NGC~253 members are
found, we compare the colour functions of the TRILEGAL simulated MW
stars to the data in different magnitude bins in
Fig.~\ref{fig:fore_cf}. For magnitudes brighter than $J \sim 20$ the
TRILEGAL MW model and the data are in very good agreement, considering
also that the synthetic diagram does not include any photometric
error, which would widen the model distributions.  In fact, the good
agreement between the TRILEGAL model and data, in
Fig. \ref{fig:fore_lf} and \ref{fig:fore_cf},  confirms that our
photometry is quite accurate for stars brighter than $J \simeq$ 20, as shown
also by the small error-bars in Fig.~\ref{fig:cmd}.
In the range $20.5 < J < 21$ the data show a strong excess of red
stars compared to the model, and some excess is possibly present also
in the range $20 < J < 20.5$. Therefore, we consider the location $J =
20.5$ as the upper limit to the apparent luminosity of the NGC~253
stellar members; below this limit the foreground contamination is
still present, but it affects mostly the blue end of the colour
distribution.   The solid green line in Fig. \ref{fig:cmd} shows our
adopted selection to isolate bona fide stellar members of NGC~253 on
the CMD. The validity of our choice is further strengthened by the cumulative
  distributions shown in Fig. \ref{fig:spatial_stars}, where stars
  brighter than $J =20.5$ closely follow the uniform distribution on
  the tile, as expected for the Milky Way foreground. 
Conversely, stars below this locus on the CMD (RGB and
  AGB) are concentrated towards the disk of NGC~253.

\begin{figure*}
\centering
\resizebox{\hsize}{!}{
\includegraphics[angle=0,clip=true]{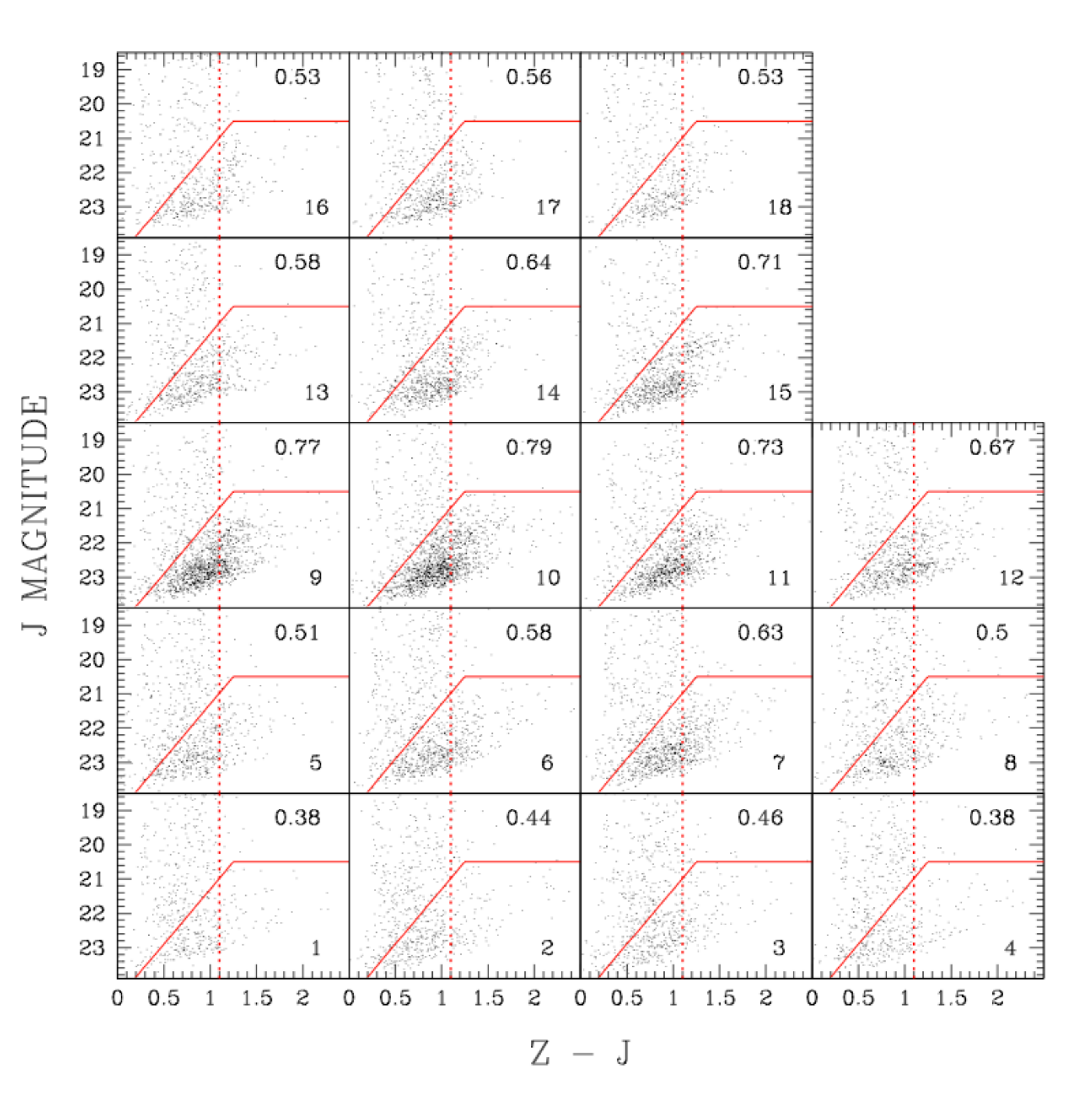}
}
\caption[]{Spatially resolved CMDs over the VISTA tile (halo
  regions). Each panel shows the CMD of the stars falling in a
  rectangular sub-region labelled in the bottom right corner with 
  the same number as in
  Fig.~\ref{fig:map}. The solid line shows the dividing line
  of the likely members of NGC~253, as in Fig. \ref{fig:cmd}.  The number
  in the upper right corner is the ratio between the stars below this
  envelope and the total number of stars in the selected region of the tile.
  Notice that the CMDs are
  plotted only up to $J=18$, while stars are measured
  up to $J=14$.  The vertical dotted line at $Z-J=1.1$ marks the
  colour at which the NGC~253 stellar population greatly outnumbers the
  foreground contamination (see Fig.\ref{fig:fore_cf}). }
\label{fig:stamps_halo}
\end{figure*}

\subsection{The CMD of NGC 253 star members}
\label{backest}

The selection criteria defined in Section~\ref{trilegal} for the
stellar population in NGC~253 can be used to verify whether there are
sub-regions in the tile where this population is scarce or absent and
derive an empirical evaluation of the contamination by foreground
stars and background unresolved galaxies.

We show in Fig.~\ref{fig:stamps_halo} the spatially resolved CMDs over
the portion of the VISTA tile which maps the halo around the galaxy,
i.e. for the rectangular sub-regions drawn on Fig.~\ref{fig:map} and
identified with progressive numbers.  The following sub-regions are excluded:
\begin{enumerate} 
\item $X<1000$ and $X>11500$ because these pixels received shorter
exposures due to the pawprint sampling;
\item the region with $7000 < Y < 11500$ is dominated by the disk of
  NGC~253, hence it is not useful to evaluate the
  foreground/background contamination;
\item we also discard the upper right corner of the VISTA tile (at
  $X>8500, Y> 11500$) because of the defects of detector 16 (see
  Sec.~\ref{sec:reduction}).
\end{enumerate}

The sub-regions targeting the halo of NGC~253 do not have the same
area because we aim at maximizing the contrast between the galaxy
population and the foreground contamination. Therefore we consider
wider areas in the outer portions of the tile.
The solid line plotted on each CMD of Fig.~\ref{fig:stamps_halo}
shows the locus below which we expect the stars members of
NGC~253 to dominate, according to the discussion in
Section~\ref{trilegal}.

 The number listed on the upper right corner of
each panel of Fig. \ref{fig:stamps_halo} is the fraction of stars in the spatially selected
sub-region which falls below the solid line, with respect to the total number
of stars in catalogue A in the same spatial region.  The vertical dotted line helps
us to evaluate the contribution from the NGC~253 members, that appear
as a red population and dominate the foreground/background
contaminants (see Fig.~\ref{fig:fore_cf}). On
Fig.~\ref{fig:stamps_halo} the number
of red stars associated with the NGC~253 population increases in
sub-regions closer to the disk, as also traced by the fraction of
stars below the solid line.

For comparison, the TRILEGAL simulation shown in
Fig. \ref{fig:fore_cmd}, has a fraction of 0.32, having taken into account
the magnitude limit drawn as a solid black line in this figure.  
This fraction is similar to the value measured on the bottom
corners of the tile, i.e. in sub-regions 1 and 4.  Although the colour
distribution in regions 1 to 4 and 16 to 18 may be consistent with
that of the foreground population, 
the excess of stars below the solid line
in regions 2, 3, 16, 17 and 18 suggests that a component of NGC 253
members is likely present in these sub-regions.  

Actually, the detailed comparison of the  
stellar population in sub-regions 1+4 with the TRILEGAL model, suitably 
re-scaled by the ratio of the sampled areas, shows a remarkable
match of the luminosity function and color distribution at $J\la 20.5$, but 
still there is an excess at faint magnitudes. 
Since the model does not include the contribution of the background unresolved
galaxies (and may have uncertain calibration at these faint magnitudes 
as mentioned above), we prefer to rely on the empirical estimate of the foreground+background
contamination using sub-regions 1+4 ($1000 \leq X \leq 3500; Y \leq
2500$ plus $8500 \leq X \leq 11500; Y \leq 2500$).
 There are 621 and
683 stellar sources in the regions 1 and 4 of the tile, respectively.  The total
number of sources is then $1304$ over a combined (region 1+4) area of $0.123$
deg$^2$ and the surface density of foreground stars included in our
catalogue of bona fide stars, i.e. our photometric catalogue A, is
empirically estimated to be $0.1\times10^5$ stars per deg$^2$,
on average. This value will be considered later, when discussing the
 extent of the NGC~253 stellar halo.

\section{NGC 253 stellar disk and halo structure}
\label{sec:structure}

Detailed analysis of the inner regions and of the disk structure in
NGC~253 is presented in Iodice et al. (in prep.). Here we discuss
primarily the structure of the halo, but to do that we need to know
how extended the disk is, and where the disk-halo transition occurs.
Since NGC~253 is not perfectly edge on, we must evaluate how far the
disk projects on the VISTA tile and extends along the projected minor
axis. We do that assuming that the disk is circular, by measuring its extension 
along the major axis and then projecting it along the minor axis using
the inclination angle of the disk.

\subsection{The stellar disk}
\label{diskstars}
\begin{figure}
\begin{center}
\resizebox{\hsize}{!}{
\includegraphics[angle=0,clip=true]{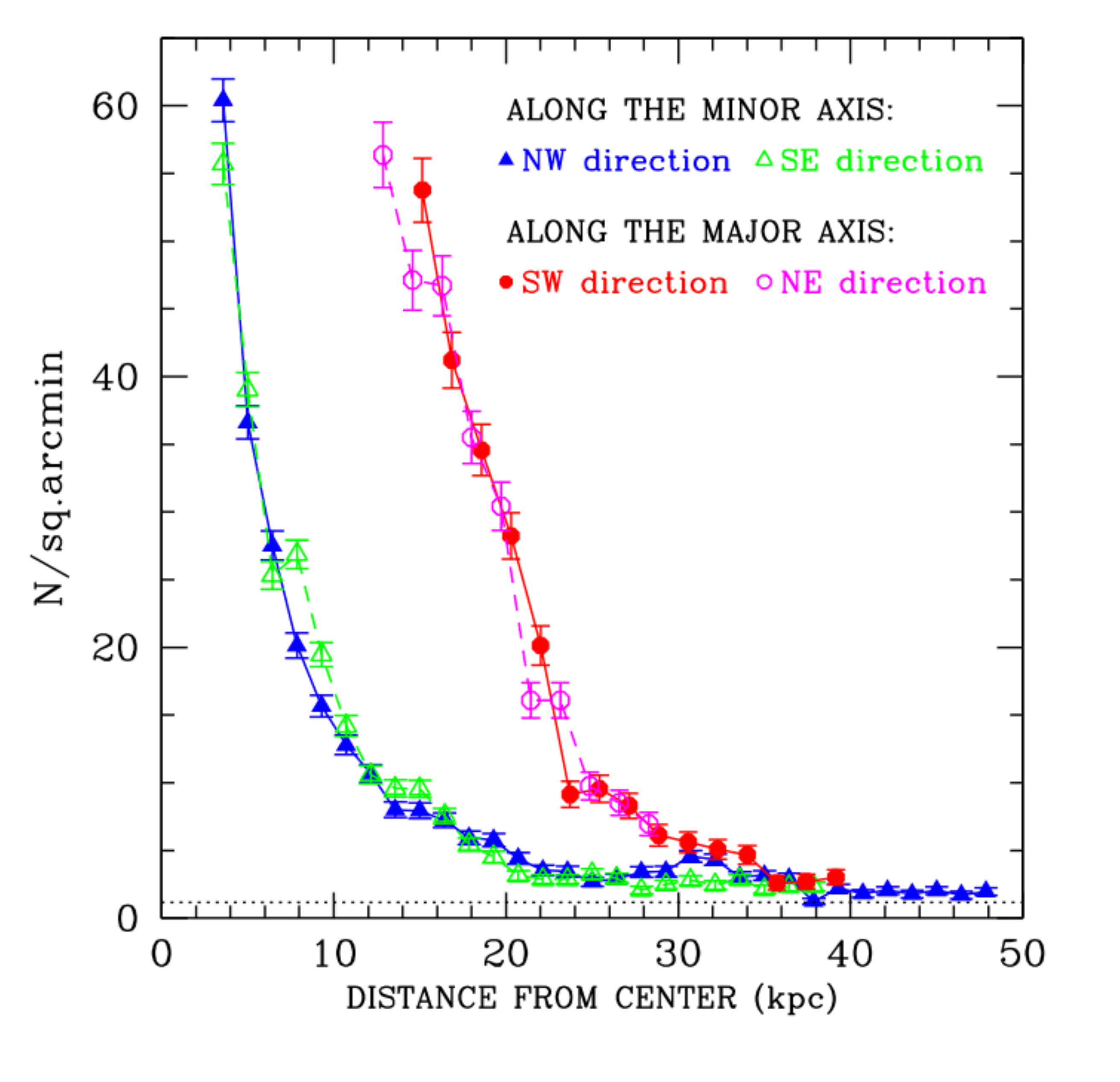}
}
\caption{Surface density profiles 
  along the major and minor axis of the NGC~253 disk, as 
  labelled. The profiles are constructed from catalogue A, considering only probable members of NGC 253 (stars below the solid line in Fig. \ref{fig:cmd}), and the error bars show the $\pm 1\sigma$ Poissonian uncertainty. The star counts along the major axis are performed within a horizontal stripe defined by 8250 $
 \leq Y \leq $ 9250 pixels, corresponding to a width of $\simeq$ 5.7 kpc. The central region is excluded due to crowding. The minor axis profiles are extracted from a stripe at 3800 $\leq X \leq$ 6800 pixels. At a distance of  $\simeq 35$ kpc from the centre the number density 
  is nearly the same in the two orthogonal directions.
 The dotted black line shows the density of foreground/background objects measured on sub-regions 1+4 in Fig. \ref{fig:stamps_halo} in the same selected part of the CMD. 
  }
\label{fig:compare_xz}
\end{center}
\end{figure}

\begin{figure}
\begin{center}
\resizebox{\hsize}{!}{
\includegraphics[angle=0,clip=true]{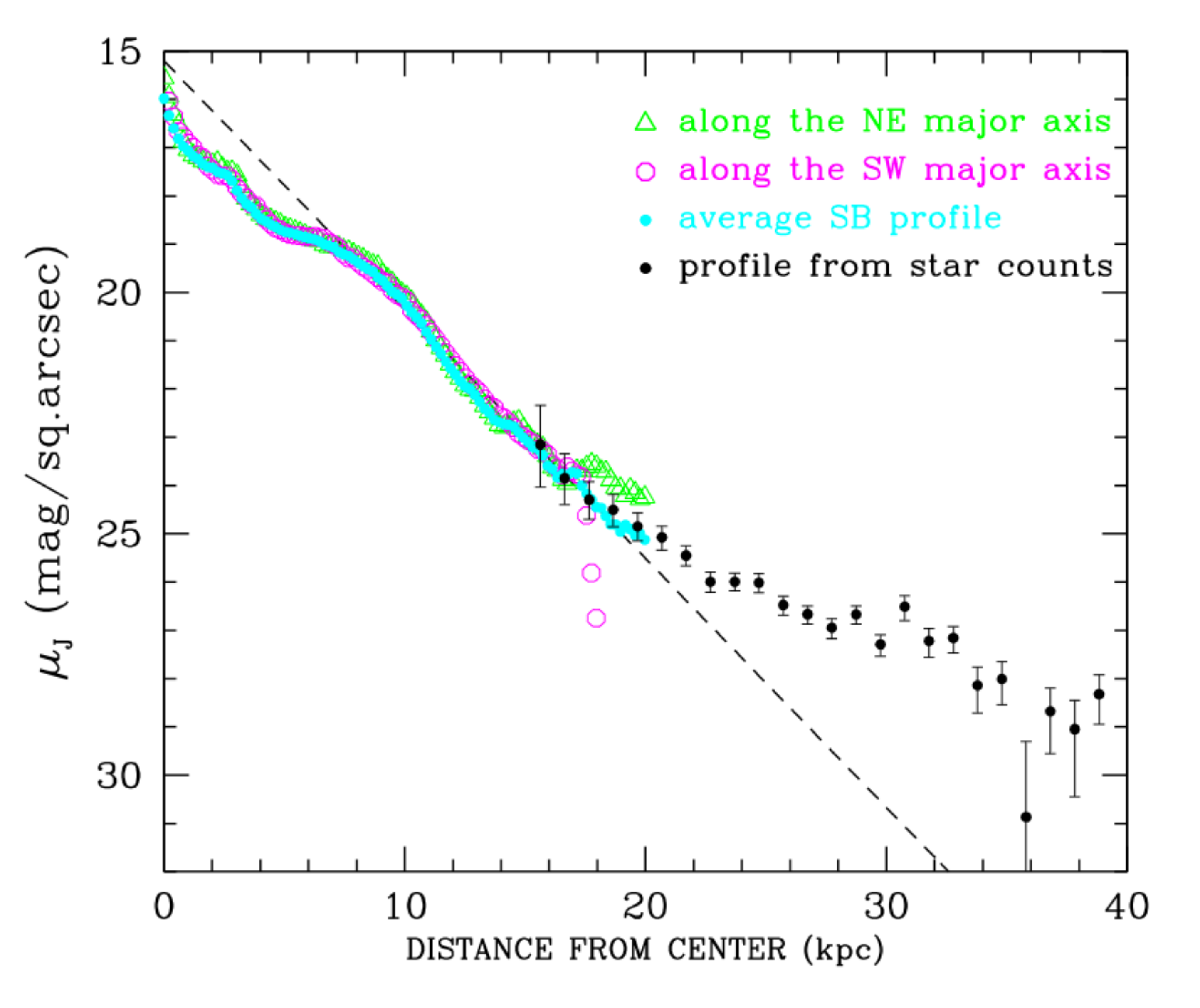}
}
\caption{Composite surface brightness profile along the major axis of
  NGC~253. The green triangles and magenta circles  show
  surface brightness profiles extracted along the North-East and South-West major axes 
  directions from the J-band early commissioning data.
  Light blue small dots trace the average J-band surface brightness profile. 
  Small black dots are the number density profile of star 
  members of NGC~253, corrected for crowding effects, and scaled to the surface brightness profile between 15-18 arcmin from the center. The error bars are Poissonian from counts  and also include the foreground/background estimated errors. The dashed line (not a fit) helps to
  better see the continuity of the surface brightness and scaled number density profiles. }
\label{fig:fit_xprofile}
\end{center}
\end{figure}

In Fig.~\ref{fig:compare_xz} we show the stellar density profile along
the two sides of the major axis (open and filled circles), 
which turn out to be quite smooth and
symmetric.  The solid line shows the counts of the NGC 253 bona fide members along the South-West direction,
while the dashed line shows the profile along the North-East direction.
The profiles from catalogue  B show very similar trends.

When the South-West and North-East semi-major axis number density
profiles are plotted on top of each other as in
Fig.~\ref{fig:compare_xz}, a change in the radial slopes appears at a
distance of approximately $25$ kpc from the centre. The South-West profile 
is more extended than the North-East one,  because the galaxy is not
centred in the tile (the disk centre is at X=5300, Y=8750 coordinate). 
We interpret the change of slope at $\sim 25$ kpc as signalling the location of the outer
edge of the disk. This interpretation is prompted also by the analysis of the surface brightness profile of the entire disk in Iodice et al. (in prep.). Since the inclination of NGC~253 is approximately
$76$ deg, the projected component of the disk along the minor axis is $\sim
6$~kpc, corresponding to $\pm 1000$ pix from the disk plane.
This implies that the regions of the tile at $Y < 7500$ and $Y >
10000$ in Fig.~\ref{fig:map}, do not include stellar populations from
the circular thin disk.  However, we know that the disk of NGC~253
appears quite disturbed, hence those regions of the tile closer to its
edge may likely include some extra-planar stellar component from the disk.

Fig.~\ref{fig:fit_xprofile} shows the complete radial profile along the major 
axis of NGC 253 out to 40 arcmin, i.e. 40.3 kpc, constructed by combining 
two surface brightness profiles (small coloured dots) extracted along North-East 
and South-West major axes 
in the $J$-band images with the number density from the $J$-band star
counts corrected for crowding effects (black dots with error bars), following the prescription
described in \citet{irwin+trimble84}.
$J$-band surface brightness profiles were extracted from an earlier
commissioning image, in which the NGC 253 disk was aligned at 90
degrees with respect to the science verification tile, hence sampling
different regions of the pawprint mosaic. These data turned out better
suited to measure the faint surface brightness in the outer regions of
the disk because of a more accurate measurement of the background. 
The average of the two $J$-band surface brightness profiles is shown
by the light blue dots in Fig.~\ref{fig:fit_xprofile}.
The surface brightness and the number density profiles were extracted
using the same projected regions relative to the major axis, a rectangular strip 4 arcmin wide beyond a
radius of 10 arcmin but tapering linearly to 1/2 arcmin at the centre,
to better sample the central regions.

\begin{figure}
\begin{center}
\resizebox{\hsize}{!}{ \includegraphics[angle=0,clip=true]{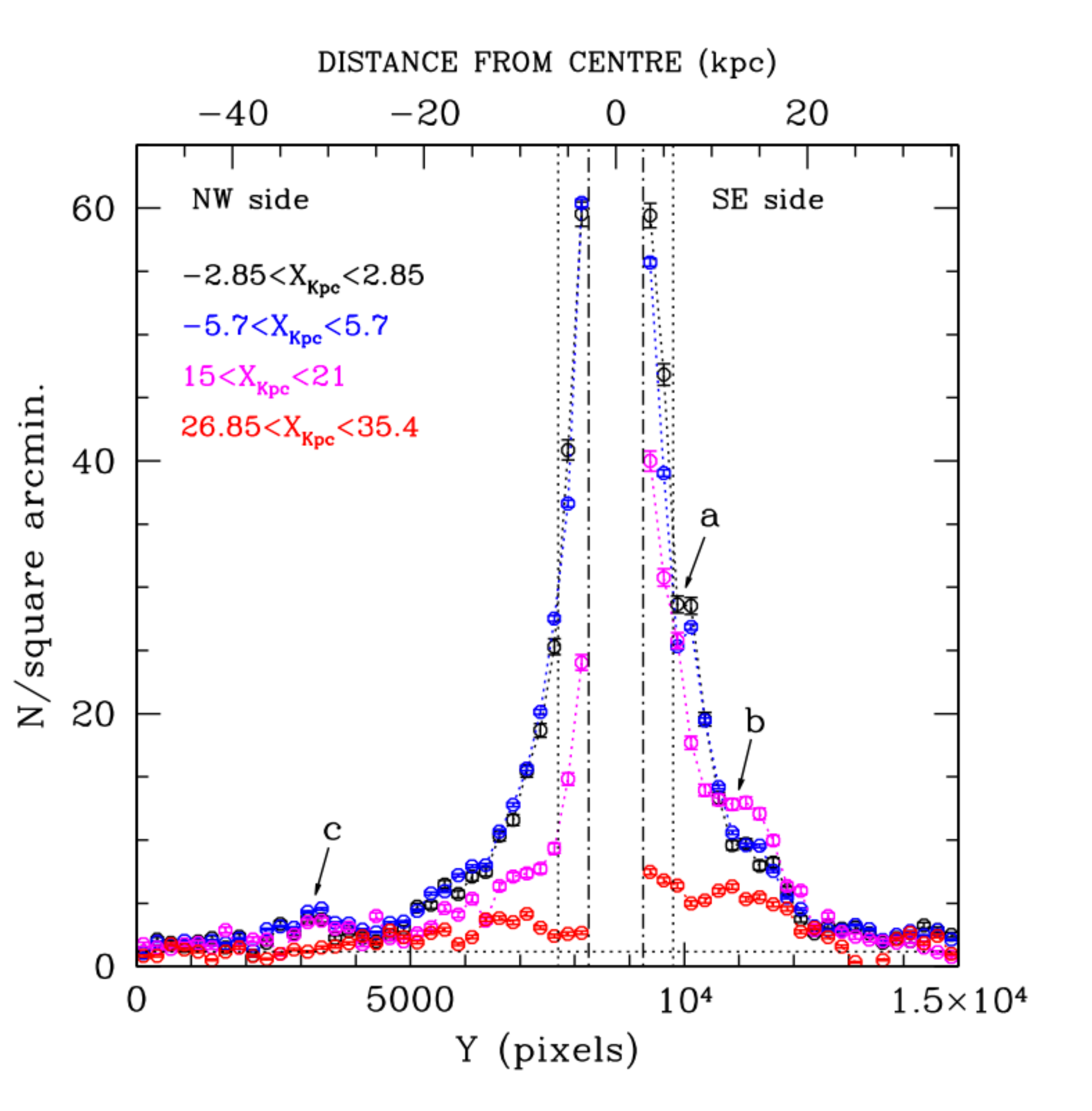} }
\caption{Surface density profiles 
  for the catalogue A likely members of NGC 253 (selected as in Fig. \ref{fig:compare_xz}) along directions
  parallel to the minor axis within four stripes as labelled.
  Dot-dashed lines limit the central region of the tile, where
  photometry is affected by very high crowding. Dotted lines are drawn
  at $\pm 6$ kpc from the galaxy major axis, showing the projected
  extension along the minor axis of the (circular) disk of the
  galaxy. Regions outside to the dotted lines should not be dominated
  by stars from the thin disk component. The horizontal line shows the
  level of the foreground/background contamination as in Fig. \ref{fig:compare_xz}.
  The arrows highlight the high
  density regions described in the text.}
\label{fig:z_profile}
\end{center}
\end{figure}

The profiles along the North-East and South-West directions are
extremely similar out to $\sim$ 16 arcmin and trace a well defined
exponential decline, with a scale length of 2.1 kpc. Beyond this
distance  the surface brightness is affected by a large error due to   
background artefacts at the level of roughly 1 ADU caused by the
complex processing. 
The profile from the star counts has been derived applying a
correction for crowding based on the seeing affecting the data so that
we could get a (small) overlapping region between the profiles computed
in these different ways. Notice that the corrected counts nicely follow the
same exponential slope as the surface brightness profile in the
innermost (between 16 and 18 arcmin) region, but 
becomes flatter at distances larger than about 20 arcmin,
($\sim$ 20 kpc), indicating the transition between the disk and the halo
component of NGC 253. The uncertainty in the sky background subtraction, which affects the surface brightness profile at $d \ga 16$ arcmin, and that on the crowding correction, which affects the counts at $d \la 18$ arcmin, implies some flexibility when matching the two components. This reflects on the slope of the matched profile and, consequently, on the positioning of the slope change. 
With our data we constrain the disk size of NGC 253 to a range between 20
(Fig. \ref{fig:fit_xprofile})  to 25 kpc (Fig. \ref{fig:compare_xz});  further analysis of the disk structure is presented in Iodice et al. (in prep.). We note that a disk size of $\sim$ 20 kpc is consistent with what measured in other galaxies \citep{Ibata+13}.

In order to match the star number counts to the surface brightness profile 
in Fig.~\ref{fig:fit_xprofile} the plotted function is:
\begin{equation}
f = -2.5\,\, \log N/A + 21.4\label{eq3}
\end{equation}
where $N/A$ is the surface density of crowding corrected stellar counts per square
arcsec. The value of $21.4$
which is needed to bring into agreement the surface photometry with star counts
can be translated into the number of detected stars per unit luminosity of the
parent population, which turns out to be $N/L_{\rm J} \simeq 10^{-4} L_{\rm
  J,\odot}^{-1}$. This number compares well with the theoretical
expectations. For the two simulations described in \citet{greggio+12}
the number of stars brighter than $M_J = -4.2$ (which
corresponds to $J$=23.5 for NGC~253) is $N_{\rm sim}/L_{\rm J} = 7 \div 9
\times 10^{-5} L_{\rm J,\odot}^{-1}$ respectively for a constant star formation rate over
the last 12 Gyr, and for an old ($\sim$ 11 Gyr) episode of star
formation.  The empirical scaling that we determine from the
comparison of the two profiles is thus in good agreement with the
predictions of stellar evolution theory.

\subsection{The halo structure and substructure}

In Fig.~\ref{fig:compare_xz} we directly compare the profiles along the
minor and major axes. Although the disk dominates most of the stripe
along the major axis, the profiles along the two orthogonal directions
seem to converge to the same level at $\sim 30$ kpc from the
centre. This is fortuitous and due to the excess counts in the
North-West profile (shown in blue) corresponding to an overdensity in the halo discussed below. Rather, the profiles along the major and
minor axes become remarkably close at distances $\gtrsim 35$ kpc from
the centre. The similarity
in the values for the number density at these distances along the two
orthogonal axes, and their excess above the foreground/background level suggests that the halo component dominates at $R> 35$
kpc and that it is nearly spherical. 

\begin{figure}
\begin{center}
\resizebox{\hsize}{!}{\includegraphics[angle=0,clip=true]{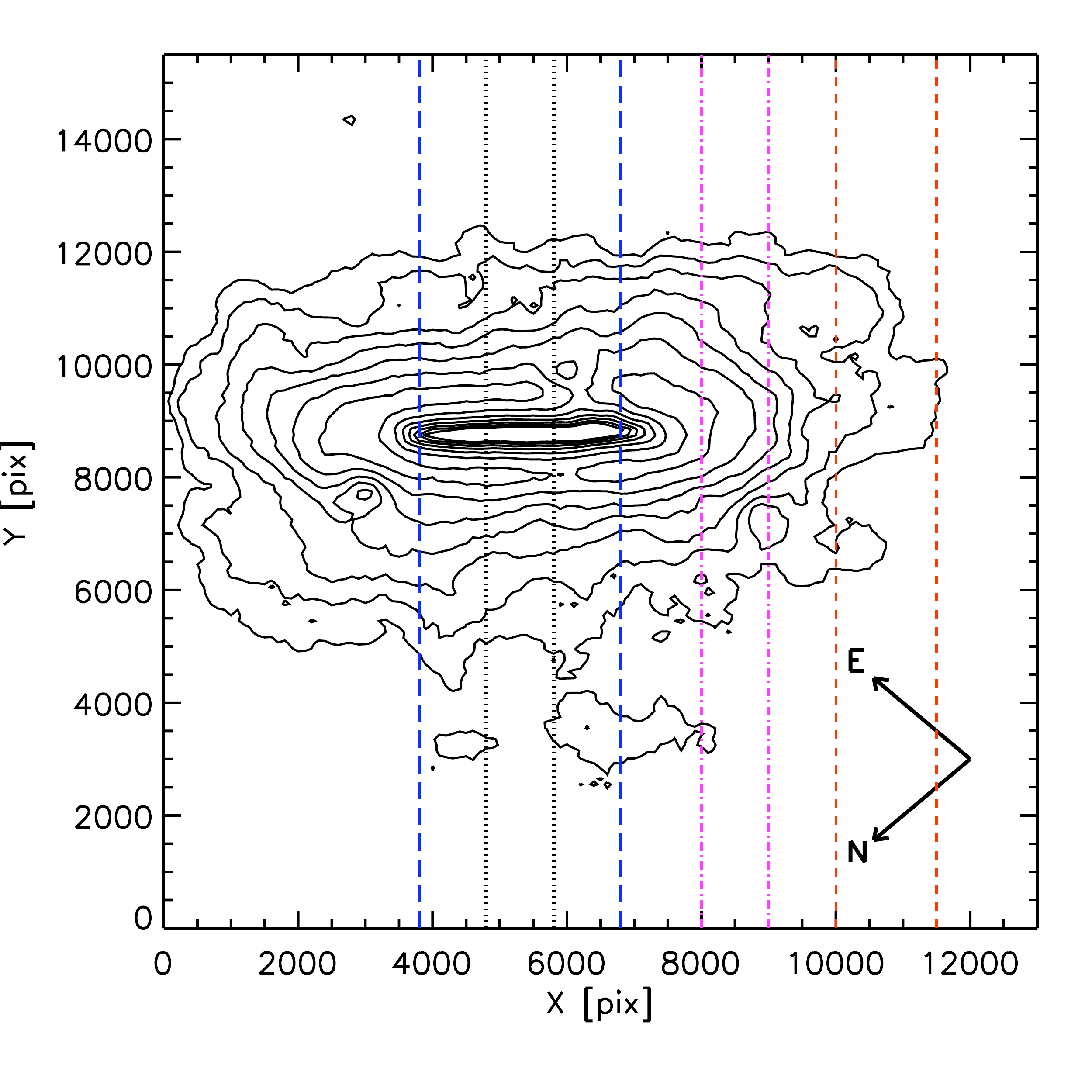}}
\caption{Isodensity contour levels for sources classified as stellar
  on both $Z$ and $J$ tiles, i.e. catalogue A. 
Contours are plotted so that, going towards the inner regions, each
contour has a 40\% higher density with respect to the previous one.
The outermost contour has
  a density of $0.23 \times 10^5$ stars/${\rm deg}^2$; the next has a
  density of $0.32 \times 10^5$ stars/${\rm deg}^2$ and so on.
  The vertical lines indicate the regions along the minor axis on which the profiles in Fig. \ref{fig:z_profile} are computed.
}
\label{fig:contours}
\end{center}
\end{figure}

In Fig.~\ref{fig:z_profile} we plot the stellar density profiles obtained from star counts 
computed in four stripes
parallel to the minor axis and indicated on Fig. \ref{fig:contours}.
The blue and black lines refer to two central stripes that include the
minor axis with two different widths. At the distance of NGC~253 they are
$5.7$ and $11.4$ kpc wide.  Notice that the profile in blue is the same minor axis profile as  in Fig. \ref{fig:compare_xz}. The steep radial decrease
of the stellar density on the central stripes does not depend on the
width of the stripe; the two profiles fall virtually one on top of
the other, indicating that the distribution is rather uniform along
the X direction.

At Y $ \simeq 10000$, i.e. at $\sim 7$ kpc from the galactic plane in
the South-East direction, the surface density in both central stripes shows an
enhancement labelled $a$, which does not have a symmetric counterpart on the
other side at $Y \sim 7500$. In Fig.~\ref{fig:map} we
identify a very bright foreground star located at $X \simeq 6000,
Y\simeq 9800$ that may be affecting the profile. However, the profile
in black, extracted from the stripe $4800 \leq X \leq 5800$, does not
include this star, and still shows the local density enhancement at
the same distance from the plane. Therefore we attribute this high
density structure to the \textit {southern shelf}
\citep{bailin+11}. The southern shelf feature is responsible for a
second overdensity  that is measured at Y $\simeq$ 11000,
both in the central stripes, and in the stripe centered at a distance of $18$
kpc from the minor axis. This second overdensity is labelled as $b$ 
and corresponds to the enhancement of stellar sources
on Fig.~\ref{fig:map} to the south of the disk. In the stripe at $18$
kpc from the minor axis, stars associated with the southern shelf can
be traced from $\sim 9$ kpc to $\sim 15$ kpc (projected distance)
above the disk.

The lowest curve (in red) on Fig.\ref{fig:z_profile} refers to a
density profile extracted from a stripe which is far from the minor
axis of the galaxy, centred at $\simeq 31$ kpc. The steep rise of the
star counts when approaching the disk of the galaxy is not detected at
this location, but we notice a density enhancement on the south-east
side that may correspond to the southern shelf. In that case the southern 
shelf overdensity is traced up to a $\sim$ 35 kpc distance from the centre of the galaxy.

On the North-West side of the tile there is a modest
overdensity at $\sim$ 30 kpc below the galaxy plane ($3000<Y<3500$),
labelled $c$ in Fig.~\ref{fig:z_profile}, that is consistently traced in three
stripes, black, blue and magenta. On the tile this looks like a filamentary
substructure located at $\sim 28$ kpc from the plane of the galaxy, 
and extending for about 20 kpc.
All these features are also clearly detected on catalogue B and this supports their reliability. 

In order to visualize the southern shelf and the overdensity at $28$
kpc below the plane of the disk we compute the contours of the density
of stellar sources from catalogue A. These isodensity contours are
computed following the method described in \citet{goldsbury10}. This
method is in general used to perform a fit of ellipses (or circles) to
the internal isocontours in order to derive an accurate centre for
stellar components. Here we use it to study the shape of the
disk and inner halo of NGC~253 and to investigate the possible
presence of overdensities. The first step is to construct a grid of
$200\times200$ resolution elements with a size of $100\times100$
pixels each. Then, the number of stars found within a circle of $200$
pixels in radius, centred on each box of the grid, corresponds to the
density level at each given element of our grid. Finally, we draw the
contours by connecting the grid boxes of equal density. 
In Fig.~\ref{fig:contours} the outermost isodensity contour has $0.23
\times 10^5$ stars per deg$^2$; we remind that the density of
the background (Milky Way foreground stars + unresolved background
galaxies) as estimated empirically in Section~\ref{backest} amounts to $0.1 \times
10^5$ stars per deg$^2$.
 
The disk of the galaxy is traced very neatly by elliptical contours,
with some irregularities due to the presence of very bright foreground
stars -- the most conspicuous being one bright star on the south east
side ($X=6075;\, Y=9790$) and two on the northwest ($X=3080,\,
Y=7750;\, X=8880,\, Y=7235$).  The \textit{southern shelf} is clearly
visible as an elongation of the contours in the upper right part of
the disk, and there seems to be a similar feature in the lower left
part of the disk: i.e. an elongated structure in the north
direction. This portion of the galaxy was marginally included in the
WIRCam images of Field 1 analyzed by \citet{davidge10}, (see
Fig.~\ref{fig:n253_fov}), but the elongation of the spatial
distribution of the stars can be appreciated only with the large field
of view and much deeper photometry of our VISTA tile. 

The contours for the number counts appear very flattened in the
innermost portion of the stellar distribution, with a projected axial
ratio of $\sim$ 0.2; this ratio corresponds to the disk inclined at 76
degrees. The axial ratio tends to increase going towards the
outer parts, but the shape remains elliptical up to the lowest density
level that we trace. For example, on the fourth outermost contour in 
Fig.~\ref{fig:contours} we estimate $b/a \simeq 0.4$, in agreement
with \citet{bailin+11}. The inner halo
is thus elliptical, as expected for the dark matter halos around galaxies
based on the cold dark matter simulations \citep{helmi04}, and similar
to what is observed for the halos of the Milky Way and Andromeda
\citep{bell+08,ibata+07}. We identify the over density at $Y = 3000$
extending from $X=6000$ to $X=8000$, which is linked to the over
density at $X \sim 4000$ at the same Y coordinate. This corresponds
to the overdensity labelled as $c$ in Fig.~\ref{fig:z_profile} and supports the
claim that we have detected a substructure at this location. 

\begin{figure}
\begin{center}
\resizebox{\hsize}{!}{\includegraphics[angle=270,clip=true]{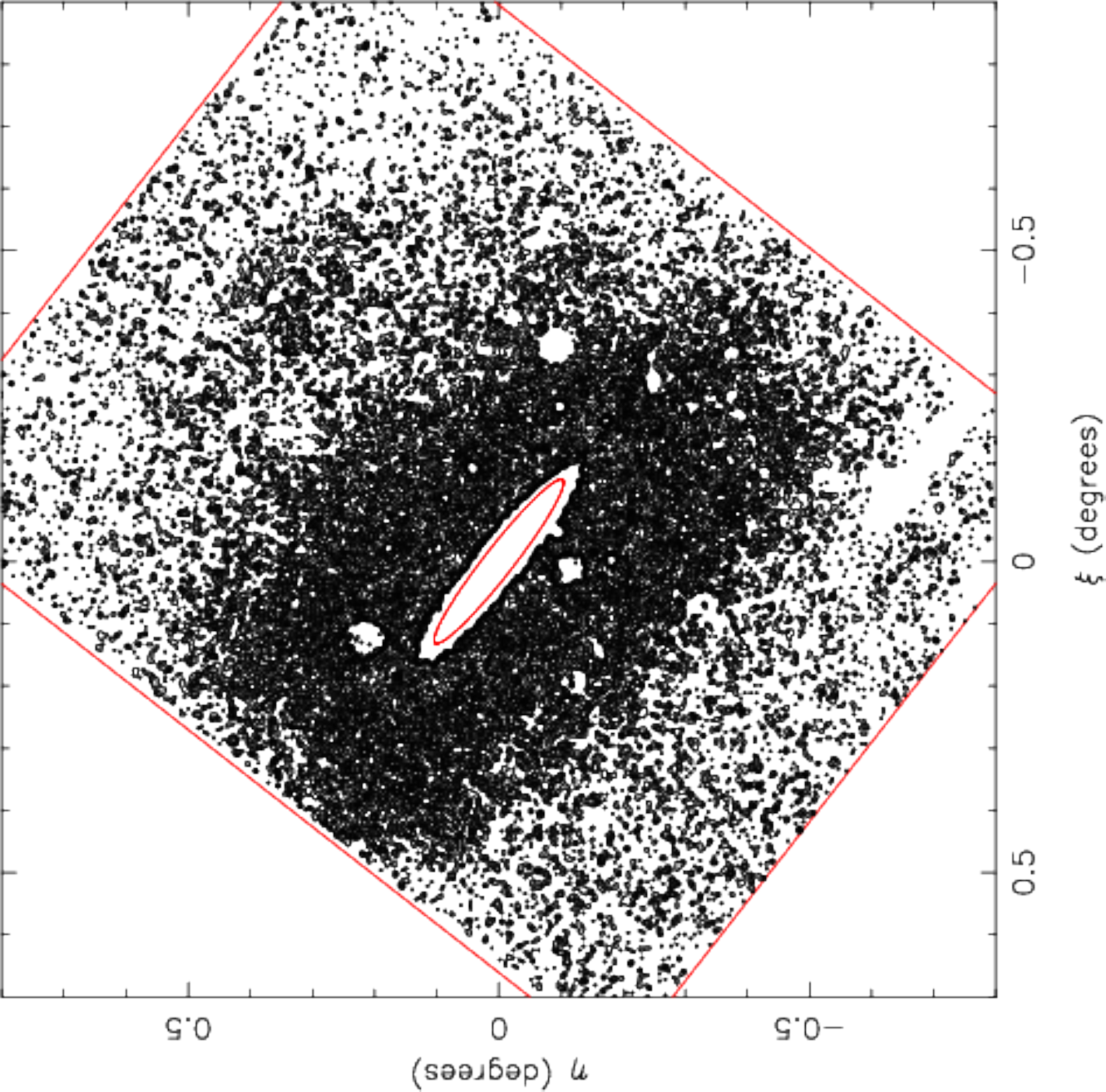}
}
\caption{Map of catalalogue A stellar like sources smoothed to a resolution of 1/2 arcmin. The orientation of the map is N to the top and E to the left and the outline
of the tile footprint is shown in red. 
}
\label{fig:cont_AB}
\end{center}
\end{figure}

This substructure in the halo is particularly evident in the map in Fig. \ref{fig:cont_AB},
which was created from a 512 x 512 grid of star counts over the tile.
The smoothing to a resolution of 1/2 arcmin is applied to highlight the
presence of the low surface brightness structure to the North-West of the main body.
Furthermore this map emphasizes the overall shape and warp/distorsion of the outer disk/halo.
Our images are not deep enough to allow a robust assessment of the
geometry of the halo beyond the outer edge in Fig.~\ref{fig:contours}, but
the matching surface density profiles along the minor and major axis beyond
$\simeq$ 35 kpc (Fig.~\ref{fig:compare_xz}) hints at a rounder outer halo.

\section{The Star Formation History}
\label{sec:SFH}

\begin{figure}
\begin{center}
\resizebox{\hsize}{!}{
\includegraphics[angle=0,clip=true]{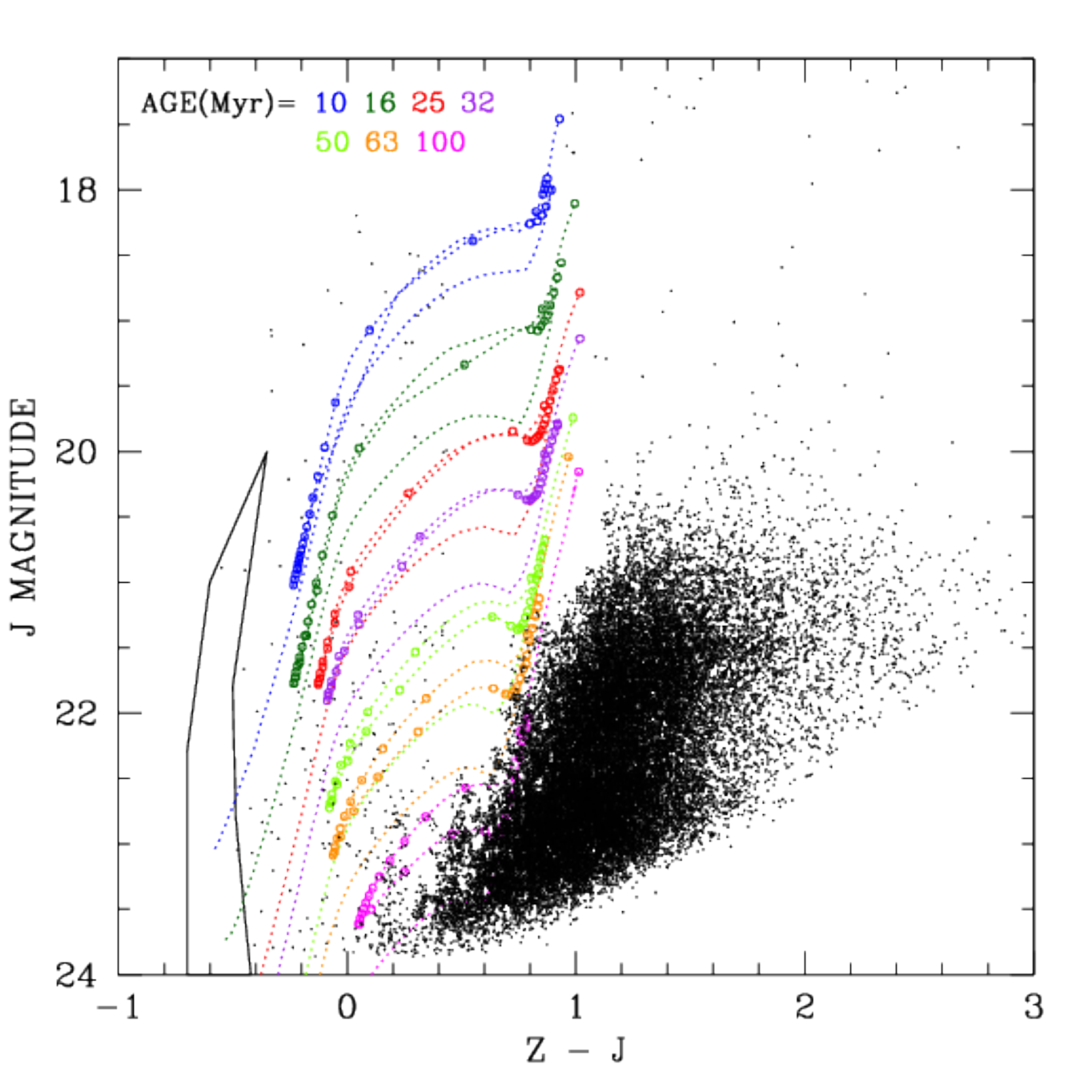}
}
\caption{Statistically decontaminated CMD for catalogue A 
  with theoretical isochrones superimposed. The solid black line limits
  the region of the main sequence (MS) stars, while the dotted lines
  connect the isochrones of the post-MS evolution for different ages,
  i.e. for 100, 63, 50, 32, 25, 16 and 10 Myr (from bottom to top).
  The open circles show the position of 30 models equally spaced in
  evolutionary mass between the turn-off mass and the mass of the
  dying star. The density of these circles is thus proportional to the
  lifetime of the evolutionary sub-phase and indicate where the stars are
  more likely located.}
\label{fig:cmd_iso_y}
\end{center}
\end{figure}

\begin{figure}
\begin{center}
\resizebox{\hsize}{!}{
\includegraphics[angle=0,clip=true]{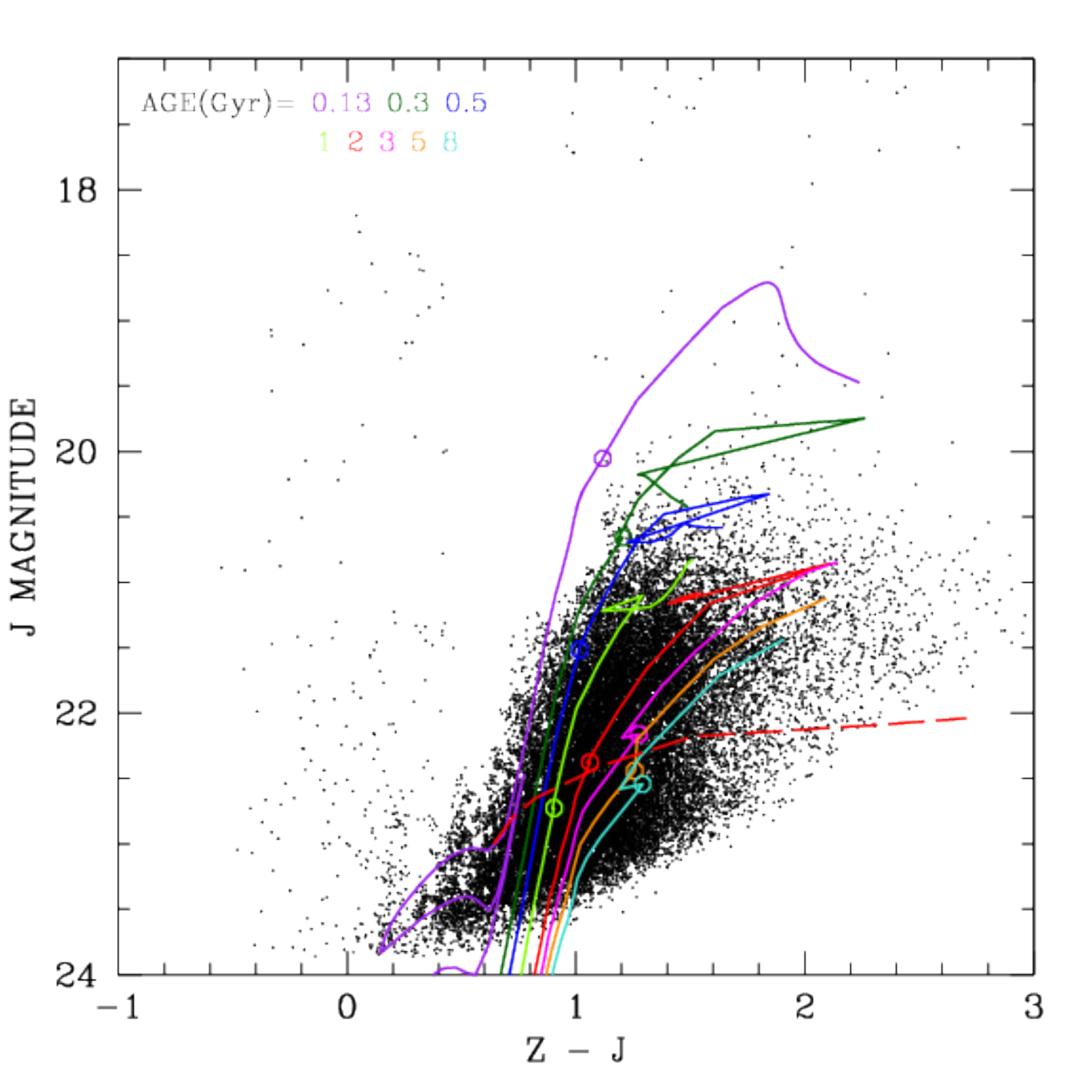}
}
\caption{Statistically decontaminated CMD for catalogue A with Z=0.008 isochrones
  superimposed. Isochrone ages are 0.13, 0.3, 0.5, 1, 2, 3, 5, and 8
  Gyr from the bluest to the reddest one. The dashed line connects the
  tip of the RGB stars of the 10 Gyr old isochrones with metallicity
  from Z=0.0001 to Z=0.03. The open circles mark the model at the
  first thermal pulse.  }
\label{fig:cmd_iso_i}
\end{center}
\end{figure}

We derive information on the star formation history of the stellar
populations in NGC~253 by comparing the observed CMD with theoretical
isocrones for different ages and metallicities. 
To do that we first clean the CMD in Fig. \ref{fig:cmd} to remove the
contribution due to foreground stars and background compact galaxies.
The statistical decontamination is performed by assuming the CMDs of regions 1 + 4 (see Fig.~\ref{fig:stamps_halo}) 
as foreground+background templates and scaling their star's
distribution by the ratio of surveyed areas. The template CMD is then
statistically subtracted from the total observed CMD by removing from
the latter the appropriate number of stars with similar colour and magnitude. 
The subtraction is done star by star, picking up 
a random star within the error ellipse, which then gets removed 
from the observed CMD \citep{zoccali+03}. 
We experimented with several 
decontamination runs, with different random number seed and slightly
different error ellipse, in order to determine the robustness of the result.
Figures~\ref{fig:cmd_iso_y} and \ref{fig:cmd_iso_i} show such 
decontaminated CMD with superimposed young and intermediate age
isochrones from the Padova database\footnote{Isochrones are computed 
using the CMD tool available from {\tt http://stev.oapd.inaf.it/cmd}}. 
The isochrones are  based on the
\citet{marigo+08} set of tracks, for a metallicity of Z=0.008.

The isochrones on Fig.~\ref{fig:cmd_iso_y} have ages 
between 10 and 100 Myr, and are plotted so that the densitity of
points indicates the sections of the isochrones where stars are 
more likely to be located. On this figure, the black line limits the
region where stars on the main sequence and with ages younger than
$\sim 15$ Myr should be found.
Since we count (almost) no stars in
this region, we conclude that there was very little (if any) star
formation in the very recent past in the outer disk and halo of
NGC~253. On-going star formation has been detected in NGC 253
\citep{engelbracht+98,comeron+01,ott+05,rs+11}, but mostly in the inner regions
of the disk, which are too crowded for our photometry.

After the statistical subtraction of foreground stars and background
compact galaxies, there are no stars left in the red supergiant
region, at ($J \lesssim 20.5$, $Z-J
\simeq 0.8$). This implies that there are (virtually) no red
helium burning stars with ages between 10 and $\sim$ 50 Myr.
Some stars (at $Z-J \lesssim 0.2$) populate the cleaned CMD in
the region of the blue portion of the core helium burning loops in
this range of ages. However, their spatial 
distribution is rather scattered, and the lack of their ``red''
counterpart further weakens the interpretation of these objects as blue
loop stars. 
Core helium burning stars with
ages between 50 and 100 Myr may be present in the CMD; those in the red section of
the loop would be found at the blue edge of the AGB/RGB tangle, those
in the blue section of the loop would be located at $J>22$ and $(Z-J) \sim
0.2$.  However, the spatial distribution of these potential blue loop
giants does not show any
concentration towards the disk of NGC 253 or association with
substructures. All stars with $Z-J \lesssim 0.2$  appear distributed
at random on the tile. We conclude that they are
most likely noise associated with the statistical subtraction of
foreground stars and background compact galaxies from the observed
CMD.

The comparison of the CMD with solar metallicity isochrones leads to
similar results; young main sequence stars are absent, as well as
giants at the blue edge of the blue loop which should populate a
vertical stripe at $Z-J \simeq 0.3$.  We conclude that star formation
was virtually quiescent over the last 100 Myr in the outer disk and
halo components of NGC~253.

Fig.~\ref{fig:cmd_iso_i} shows the comparison between the
statistically decontaminated CMD and isochrones with ages between 0.13
and 8 Gyrs, which encompass well the distribution of our data. 
These isochrones come from the same dataset as those in
Fig.~\ref{fig:cmd_iso_y}, and have a metallicity of Z=0.008. We
considered lower and higher metallicity models, but found that 
these provide a worse match to the measured colours, the former being
too blue and the latter too red with respect to the stars sampled by
our CMD. The dashed line in Fig.~\ref{fig:cmd_iso_y} shows the
position of the tip of the RGB for a wide range of metallicities (i.e. the
same as the dotted line in Fig.~\ref{fig:cmd}).  The stars brighter
than the RGB tip, but fainter than $J \sim 20$, are in the AGB
evolutionary phase. Comparing the position of the first
thermal pulse on each isochrone (indicated with a circle) to the
data we conclude that most of the AGB stars on our CMD appear to be in the
thermally pulsing phase. Early AGB stars may be found at the blue end
of this distribution, in which case they may belong to a relatively
young component. Alternatively, the blue edge of the distribution
could include relatively older stars scattered to the blue by
photometric errors.

There are very few stars, if any, around the
youngest of the plotted isochrones in its bright section. If a
conspicuous star formation episode took place between 0.1 and 0.3 Gyr ago we would
sample its AGB progeny at $J \lesssim 20$, which is not there. The
upper limit of $J=20.5$ to the NGC 253 members, discussed in the
previous section, translates into a lower limit of 0.5 Gyr for the age
of the most recent star formation activity in the region of the galaxy
that we are investigating.  Notice, however, that since some bright stars 
between the $0.5$ and $0.3$ Gyr isochrones are present in our CMD, 
the lower limit to
the stellar ages is estimated between $0.5$ and $0.3$ Gyr in our
surveyed region. The bright AGB population extends to very red
colours, in principle consistent with ages as old as 10 Gyr, for
Z=0.008.  The red bright AGB stars, however, could also have higher
metallicity and younger ages.

\begin{figure*}
\centering
\resizebox{\hsize}{!}{
\includegraphics[angle=0,clip=true]{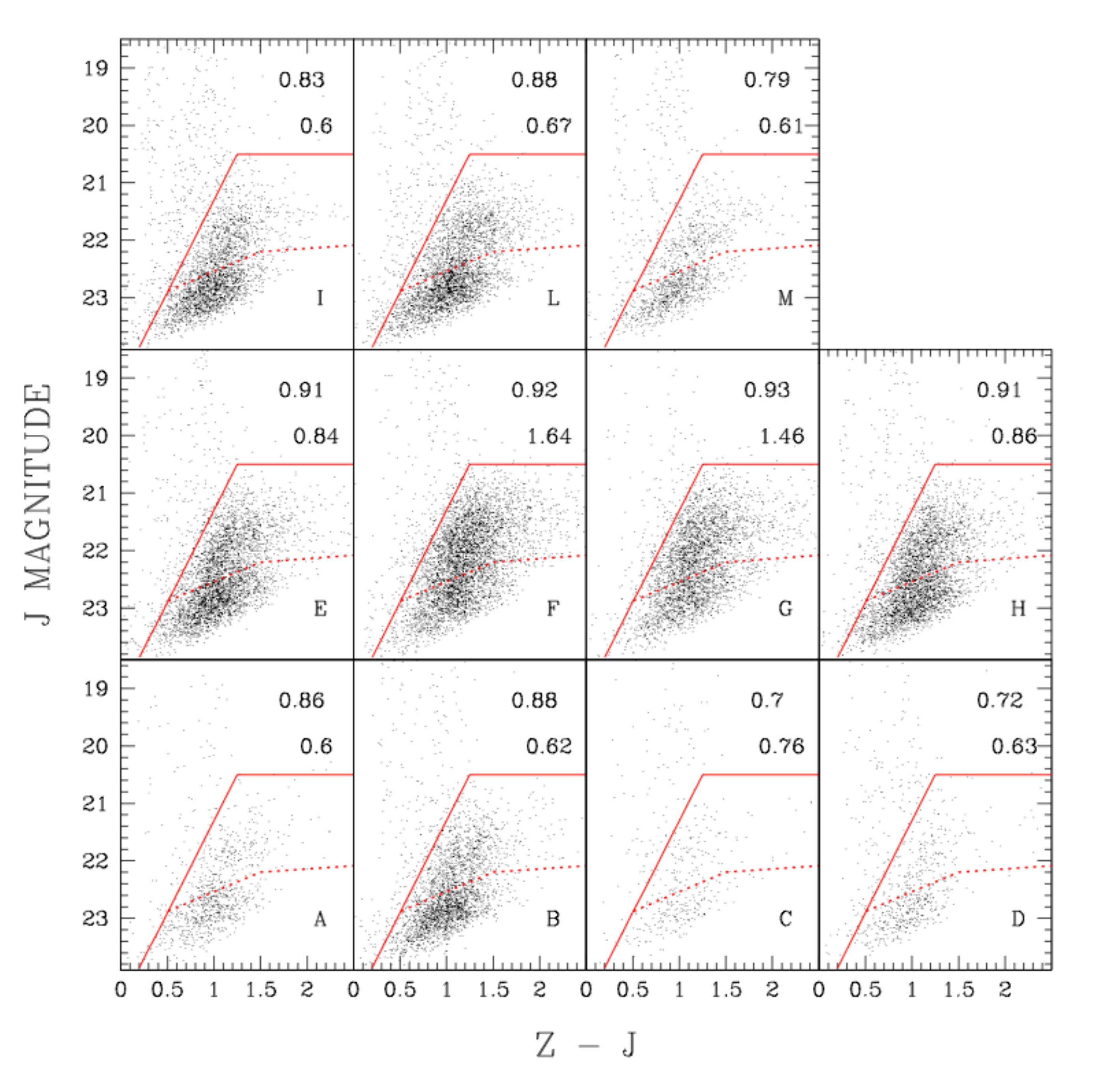}
}
\caption{CMD of catalogue A stars in sub-regions of the tile covering
  the outer disk of NGC~253, and shown on Fig.~\ref{fig:map}. The CMDs
  are identified with the same alphabetic letter as the sub-regions in
  the outer disk in Fig.~\ref{fig:map}.  The solid line shows
  the selection criteria for star members of NGC~253; the dotted
  line shows the location of the tip of the RGB stars as a function of
  metallicity as in Fig.~\ref{fig:cmd}.  The two numbers listed in
  the upper right corner of each panel indicate (uppermost) the
  fraction of NGC~253 star members over the total number of stars
  detected in the sub-region (i.e. the same as reported in
  Fig.~\ref{fig:stamps_halo}), and (lowermost) the ratio between the bright
  AGB stars (counted between the solid and dotted lines) and the
  RGB stars (counted below the dotted line). The fraction of AGB
  stars is a function of position in the tile and it increases as we
  come closer to the disk of NGC~253. }
\label{fig:stamps_disc}
\end{figure*}

The number of bright AGB stars per unit mass of the parent population,
hereafter the specific production of bright AGB stars ($P_{\rm AGB}$), is 
large at intermediate ages, but becomes small at old
ages, due to the action of mass loss on the relatively small 
envelope of low mass thermally pulsing AGB stars
\citep{noel+13,greggio+renzini11}. Using \citet{marigo+08} models, we
compute $P_{\rm AGB}$ in the
region of the CMD brighter than the (red) dashed line on
Fig.~\ref{fig:cmd_iso_i}  from stellar populations with ages between
0.3 and 3 Gyrs. Assuming a  Salpeter-diet IMF (i.e. $\phi(m) \propto
m^{-1.3}$ between $0.1$ and $0.5$ \msun; $\phi(m) \propto m^{-2.35}$
above $0.5$ \msun), models with $Z=0.008$ yield $P_{\rm AGB}$=
(0.8, 1.7, 1.4, 0.3) $\times 10^{-4} \msun^{-1}$ respectively for ages
of (0.3, 0.6, 1, 3) Gyr. The mean value of the specific production
over this age range turns out to be 0.8 
stars per $10000 $\msun.
Since we count $\sim$ 17000 stars in this part 
of the CMD, the AGB bright component traces about $2 \times
10^8$ \msun\ of stars formed between $0.3$ and $3.0$ Gyrs ago, 
if the star formation rate was constant between these epochs. 
The stellar mass formed at intermediate ages 
would be different by a factor of $\sim$ 2 if the star formation was
peaked at a specific age. However,
the color distribution of bright AGB stars does not support a bursting
star formation history. An additional uncertainty
in this mass estimate comes from systematics in the models and assumed
IMF and metallicity. The evaluation of these effects is beyond the
scope of this paper.
Most of this star formation occurred in the disk rather than in the halo of NGC~253.
We count $\simeq 12500$ bright AGB members of NGC 253 in the disk (specifically
within $1000 \leq X \leq 9500$ and $7000 \leq Y \leq 10500$), which is
close to $75 \%$ of the total detected on the tile.

Figure~\ref{fig:stamps_disc} shows the spatially resolved CMD for the
section of the tile covering the outer disk of NGC~253. 
The sub-regions, identified with the same letter in the lower right
corners as in Fig.~\ref{fig:map},
have been designed with the aim of exploring systematic variations of
the CMD morphology in different parts of the galaxy. Regions A and
B sample the more external portion of the disk, while regions C and D 
also probe the inner halo. Regions E, F, G and H target the disk; region L
is centred on the southern shelf, while regions I and M sample the
inner halo on two opposite sides of region L. 
The overall appearance of the CMD is quite homogeneous over the outer disk, with
similar upper boundary and colour range covered by the stellar members of
NGC~253.   

 The upper right corner of each panel reports two numbers. The upper
 number is the analog of the fraction reported in the upper right
 corner on each panel of Fig.~\ref{fig:stamps_halo}, i.e. the fraction of NGC~253 star members
over the total number of stars detected in the sub-region.
This fraction increases as one approaches the disk of NGC~253, and nicely complements
the trend shown in Fig.~\ref{fig:stamps_halo}. 
 The lower number is the ratio between the bright AGB stars, counted
between the solid and dotted lines, and the RGB stars, 
counted below the dotted lines.  This fraction increases markedly
towards the disk on NGC~253, so that in regions F and G the bright AGB
component outnumbers the faint RGB one. Part of this trend is due to
the effect of crowding which makes stars appear brighter than they are
\citep[see e.g.\ ][]{greggio+renzini11} and at the same time hampers
detection of the faint RGB component. 

The shift in the luminosity function towards the bright magnitudes
artificially induced by crowding may also be responsible for the
higher fraction in panels F and G of bright AGB stars close to the
upper envelope (solid line). On the other hand,
both the higher AGB-to-RGB ratio and the more populous upper envelope of
the AGB stars could indicate a stronger star
formation between 0.3 and 0.5 Gyr ago in these parts of the disk with
respect to those further out. It is also plausible that the blue
portion of the AGB region of the CMDs in the inner disk includes some
red supergiants (see Fig.~\ref{fig:cmd_iso_y}). This younger population
is recognized in the higher spatial resolution HST data \citep{rs+11},
especially in the more central parts.
Actually, the detailed study of the star formation history in the
disk of NGC~253 can be best performed using the HST data.

Similar considerations hold for the regions of the outer disk shown in
Fig.~\ref{fig:stamps_disc}, i.e. we cannot state with confidence
whether the observed gradient of the ratio between the bright AGB and
the RGB stars reflects a real gradient of star formation, with regions
B, I, L and M less active in the more recent past with respect to
regions E and H, and the latter less active than regions F and G.  As
mentioned above, the morphology of the CMDs appears quite homogeneous,
indicating similar star formation histories. Nevertheless we remark on
the following differences: in region I there is a horizontal feature
at $J \simeq 21.5$ extending to very red colours, reminiscent of carbon
giant stars \citep[see also] []{davidge10}. 
A similar morphology can be traced on the CMD of regions
L and M, but is much less evident on the CMDs of the other
regions. The CMD of region L, which is centred on the
\textit{southern shelf}, shows a wider AGB with respect to the two adjacent
regions. Conversely, the AGB spans a relatively narrow colour range in
regions E and B. These variations in morphology may indicate different
metallicity distributions, and details of the star formation history;
a quantitative analysis of the population gradient needs the
computation of synthetic CMDs taking into account the detailed
photometric quality of our data, which is beyond the scope of this work.

\begin{figure}
\begin{center}
\resizebox{\hsize}{!}{
\includegraphics[angle=0,clip=true]{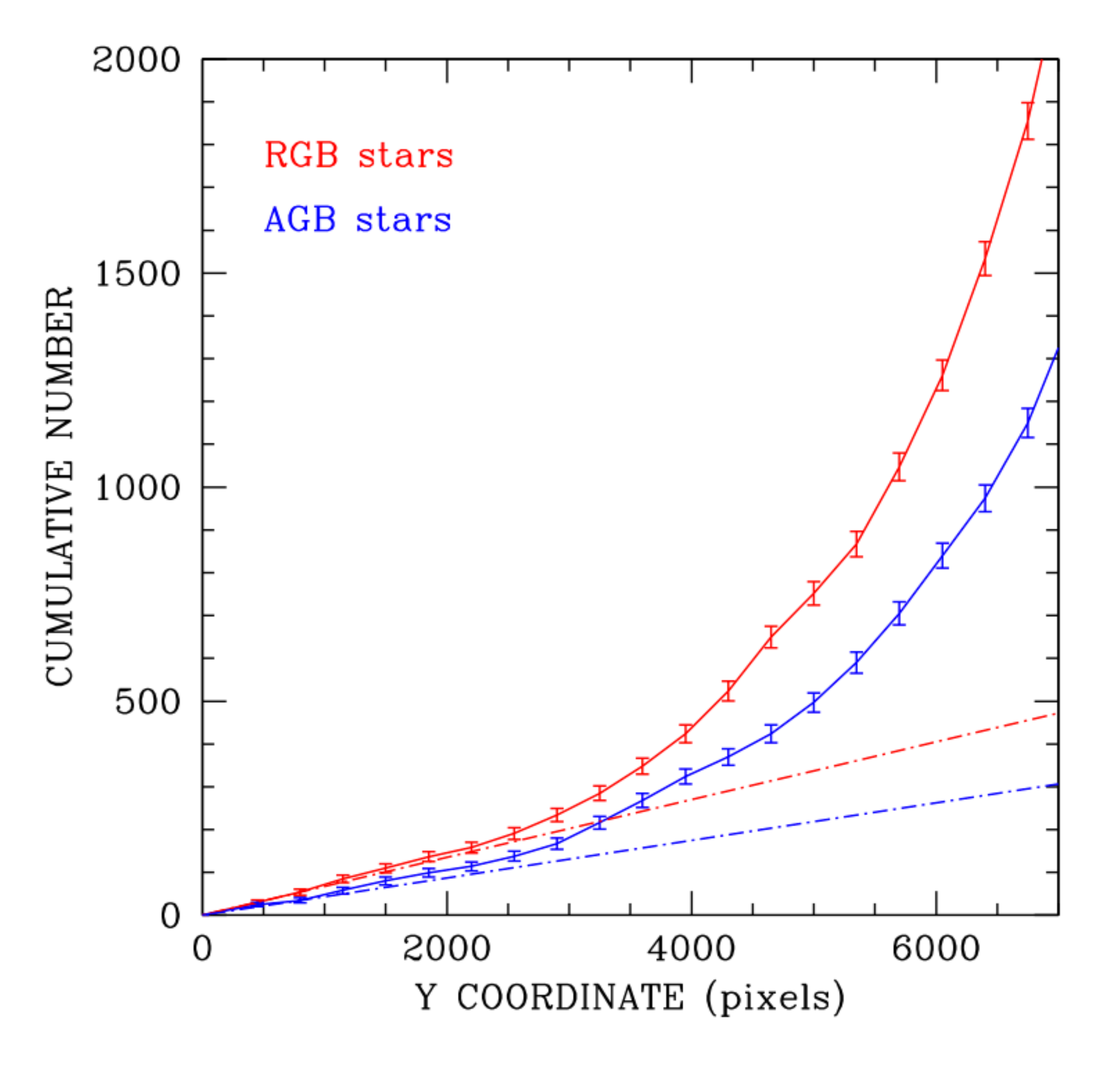}
}
\caption{Solid lines: distribution of the RGB (upper) and AGB (lower) stars extracted from the statistically decontaminated catalogue and located in the central portion of the tile ($2500 \leq X \leq 8500$). The error bars show the $\pm 1\sigma$ Poissonian uncertainty.  Dash dotted lines: straight lines fitting the distribution up to $Y=1000$.  }
\label{fig:halo_extent}
\end{center}
\end{figure}

\section{Summary and Discussion}\label{sec:discussion}

We present the results based on deep $Z$ and $J$ band VISTA photometry
of the spiral galaxy NGC~253 extending up to
$\sim$ 40 kpc from the centre along the major axis (in the South-West
direction) and up to $\sim$ 50 kpc along the minor
axis (in the North-West direction), which enables the
study of the stellar population associated with the outer disk and halo
of this spiral galaxy.

Adopting the distance modulus derived by \citet{rs+11} from ACS
observations, we are able to place the tip of the RGB at $J \simeq$
22.5 and determine the nature of the NGC~253 members visible in our CMD plots.
These appear to consist of bright RGB stars and AGB stars. The excess number counts in
the CMD over the empirically estimated foreground/background trace halo star members of
NGC~253 as far as $50$ kpc from the centre. The spatial distribution of these excess counts leaves little room for an alternative explanation, as can be appreciated from 
Fig. \ref{fig:halo_extent}, since  the cumulative counts start deviating from a straight line at small 
$Y$ coordinate. The uncertainty on the level of the foreground/background contamination, coupled with the relatively bright limiting magnitude of our data, prevent a robust determination of the size of the halo, but several evidences suggest that it extends over the whole tile and possibly beyond (Fig. \ref{fig:stamps_halo},  Fig. \ref{fig:compare_xz}, Fig. \ref{fig:halo_extent}).  Counts of sources in catalogue B confirm these findings.

The profile along the whole major axis is plotted on Fig. \ref{fig:fit_xprofile}
which illustrates the complementarity of the surface brightness and number counts methods to evaluate the light profile on the disk. Individual
star counts become very effective in those regions where the sky
subtraction and data reduction systematics become the dominant source of uncertainty for the direct determination of the surface brightness.
On the profile we clearly detect a break  with a net change of slope, that we interpret as
the location of the outer edge of the disk (see also Iodice et al., in prep.).
The coordinate at which this break occurs is sensitive to several uncertain factors, among which the most important are the estimate of the sky contribution, and the incompleteness correction applied to the data. As a result,  the profile fitting is not well constrained and we place the transition between the disk and the inner halo component at a distance between 20 and 25 kpc from the center, the latter value supported by the uncorrected counts in Fig. \ref{fig:compare_xz}. We remark that the change of slope is detected in the star counts, the surface brightness being affected by a too large uncertainty at such distances from the center.
Given the inclination of the NGC~253 galactic plane, we infer then that the thin circular disk is
confined to the very central part of the tile (at $7500 \lesssim Y
\lesssim 10000$). 

The disk appears disturbed, presenting extraplanar components, i.e. the well known \textit{southern shelf} and a substructure in the North which implies an almost symmetric North-Sourth warp in the outer contours of the disk (see Figs. \ref{fig:map}, \ref{fig:contours}, \ref{fig:cont_AB}). This kind of feature suggests that an event of  strong disturbance happened in the past, possibly the merging with a relatively massive galaxy \citep{davidge10}. We remark that thanks to the very wide VISTA area surveyed the shape of the disk is very well traced, in particular the 
northern disturbance is revealed for the first time.

In addition, we detect for the first time a new substructure in the halo of the galaxy, North-West of the main disk. This substructure is evident in the stellar density profiles along the minor axis (Fig.~\ref{fig:z_profile}) as well as in the contour maps (Figs.~\ref{fig:contours} and \ref{fig:cont_AB}). It is located about $28$ kpc from the plane of the galaxy along its minor axis, and extends for about $20$ kpc. Likely this is a stellar stream remnant of a previous accretion
episode. 
Our photometry only probes the brighter portion of the RGB; this
corresponds to a lower limit to the mass of the stellar populations
which can be revealed in our counts. From theoretical evolutionary models
\citep{marigo+08}, and assuming a Salpeter-diet IMF as in Section~\ref{sec:SFH}, 
the specific production (i.e. number of stars per unit mass of the
parent population) in the upper magnitude bin of the RGB ranges between $\sim (9  \div 3) \times 10^{-5} \msun^{-1}$  as the age
goes from 3 to 10 Gyr, with little dependence on the metallicity.  
As a consequence, our star counts trace only relatively
massive underlying populations, i.e. on the average we count 1 bright
RGB star per 20000 \msun\ of parent stellar population. Given our
background contamination of $10^4$ stars per 
deg$^2$, we estimate a lower limit to the mass surface density
of detectable enhancements due to the NGC~253 stellar members of 
$2 \times 10^{8} \msun$/ deg$^2$ corresponding to $15
\msun$/arcsecond$^2$,  or 0.053 $\msun/{\rm pc}^2$ at the distance of NGC~253.
A less massive and/or a more disperse stellar population goes
undetected below the general background. The substructure mentioned
above has a surface density more than twice the background. We
estimate the mass of the parent stellar population by determining the
excess star counts in the region defined as ($3000 \leq Y \leq 3500$;
$1000 \leq X \leq 10000$). In this region we count 780 stars, where, given
the area of 0.04 deg$^2$, we expect a background of 400 objects.
The excess of 380 counts corresponds to a mass of the parent stellar
population of $\sim 7.5 \times 10^6 \msun$. This is an approximate
estimate with several factors contributing to its uncertainty. For
example, our data suffer from some incompleteness especially at the
faint end, so that we could have underestimated the number of bright RGB stars. 
On the theoretical side, the specific production of
bright RGB stars depends on the age and metallicity of the parent stellar
population, and our adopted average value could be
inaccurate, depending on the actual star formation history of this
component. Nevertheless, the above estimate is expected to be
correct within a factor of a few  (for the assumed IMF), and we speculate
that the filamentary structure detected in this region of the halo of NGC~253 is
the remnant of an accreted  low mass dwarf galaxy, or the nuclear
part of a more massive object. 

\begin{figure}
\begin{center}
\resizebox{\hsize}{!}{
\includegraphics[angle=0,clip=true]{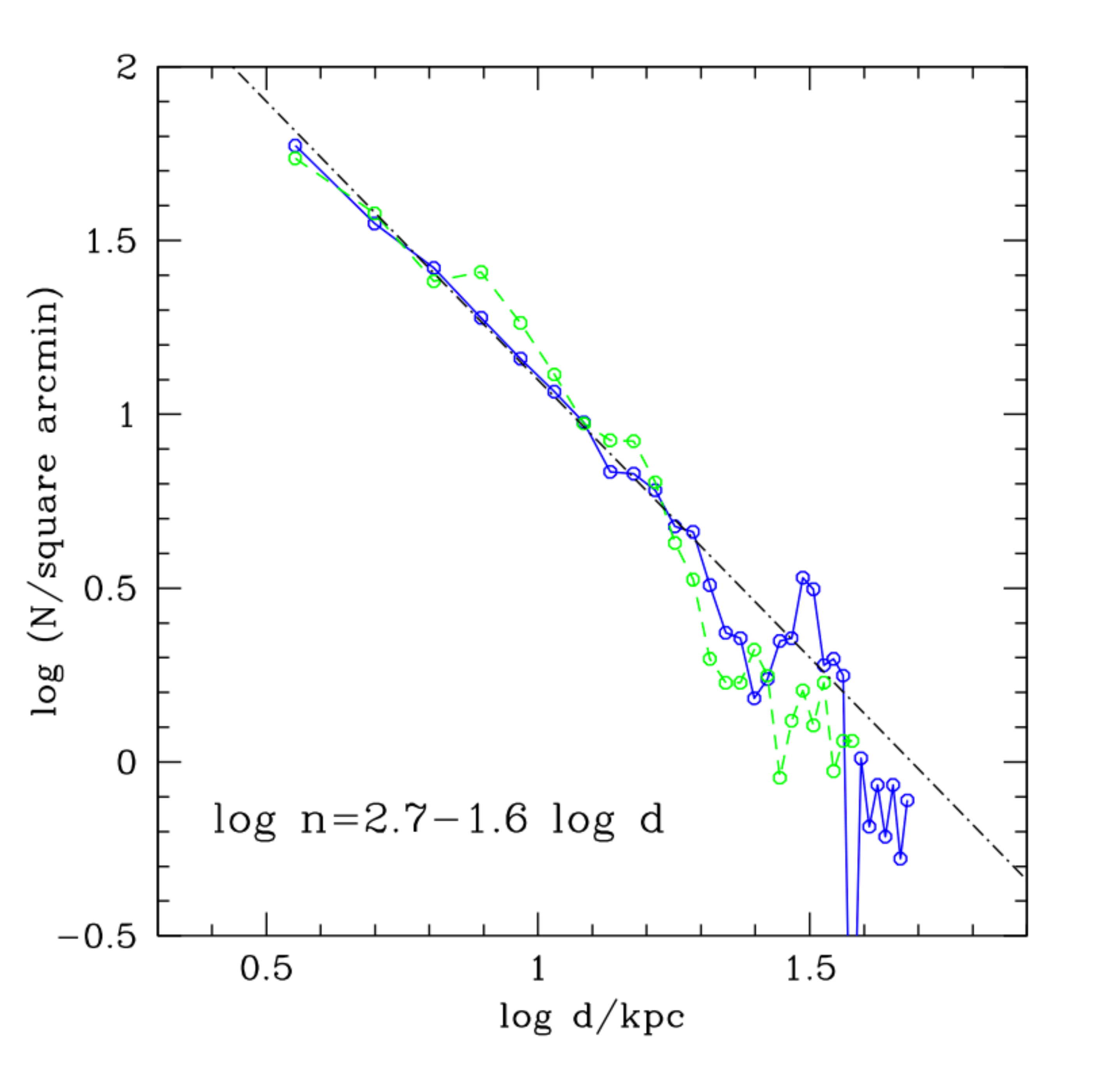}
}
\caption{Surface density profiles of NGC~253 member stars along the
  minor axis of the disk on a log-log scale. The blue solid line traces the density
  along the North-West and the green dashed line traces the
  South East direction. The empirical foreground/background contamination has been subtracted from the counts. The dot-dashed line shows the relation labelled.}
\label{fig:z_log_prof}
\end{center}
\end{figure}

The inner halo appears of elliptical shape
close to the disk and tends to become rounder further out, as indicated by  the merging of major and minor axis profiles at $\sim$ 35 kpc (Fig. \ref{fig:compare_xz}), and the excess star counts in Regions 2 and 3 compared to Regions 1 and 4 in Fig. \ref{fig:stamps_halo}.
Our images are not deep enough to allow a robust assessment of the
geometry of the halo beyond the outer edge in Fig. \ref{fig:contours}. 
Deeper data, yielding a larger number of RGB stars, would have
enhanced the statistical significance of the shape of our contours,
and lessened the effect of the bright foreground objects.

Fig. \ref{fig:z_log_prof} shows  the profile of the AGB and RGB stars belonging to NGC~253 
 along the minor axis computed on a central stripe 11.4 kpc
 wide (the area between the blue long dashed lines in Fig. \ref{fig:contours}), on a logarithmic scale. The counts in the fiducial region of the CMD have been corrected for the empirical foreground/background contamination of $\simeq 1.1$ stars per sq.arcmin. The peak
at $\sim$ 30 kpc  corresponds to the substructure discussed above.
On average, the two sides of the profile are extremely similar, and, up to 20 kpc from the galaxy plane, very  well represented by a power law with a slope of $n=-1.6$, equivalent to a
slope of $-4$ for the surface brightness profile, when expressed in 
mag per square arcsec. Beyond 20 kpc the profile seems to become steeper, but the presence of the overdensity at 28 kpc  complicates the description of its shape. In addition, we remark that this steepening is much less evident if the counts are not corrected for the foreground/background contamination.  Since this correction is uncertain we cannot conclude  on the significance of this steepening. The slope of the power law which well describes the inner portion of the halo
is not far from what is measured for other spiral galaxies. 
For M33 \citet{grossi+11} find $n \simeq -2.1$; for M31 
\citet{ibata+07} determines $n=-1.9$, while \citet{tanaka+10} find
$n=-2.2$ and \citet{courteau+11} an even steeper slope of $n=-2.5$.
The reason for the different values obtained fitting the same galaxy
is not clear. Theoretical models by \citet{font+11} predict an average
slope of $n =-2.5$  (in the same units) but with large case to case
variation. Interestingly,  when normalized to the
inner region (within 30 kpc), the slope in \citet{font+11} models is shallower, and values as low as
$n=-1.6$ can be found. We conclude that our determined slope is in a
broad agreement with the models, although relatively flat. 

\begin{figure*}
\centering
\resizebox{\hsize}{!}{
\includegraphics[angle=270,clip=true]{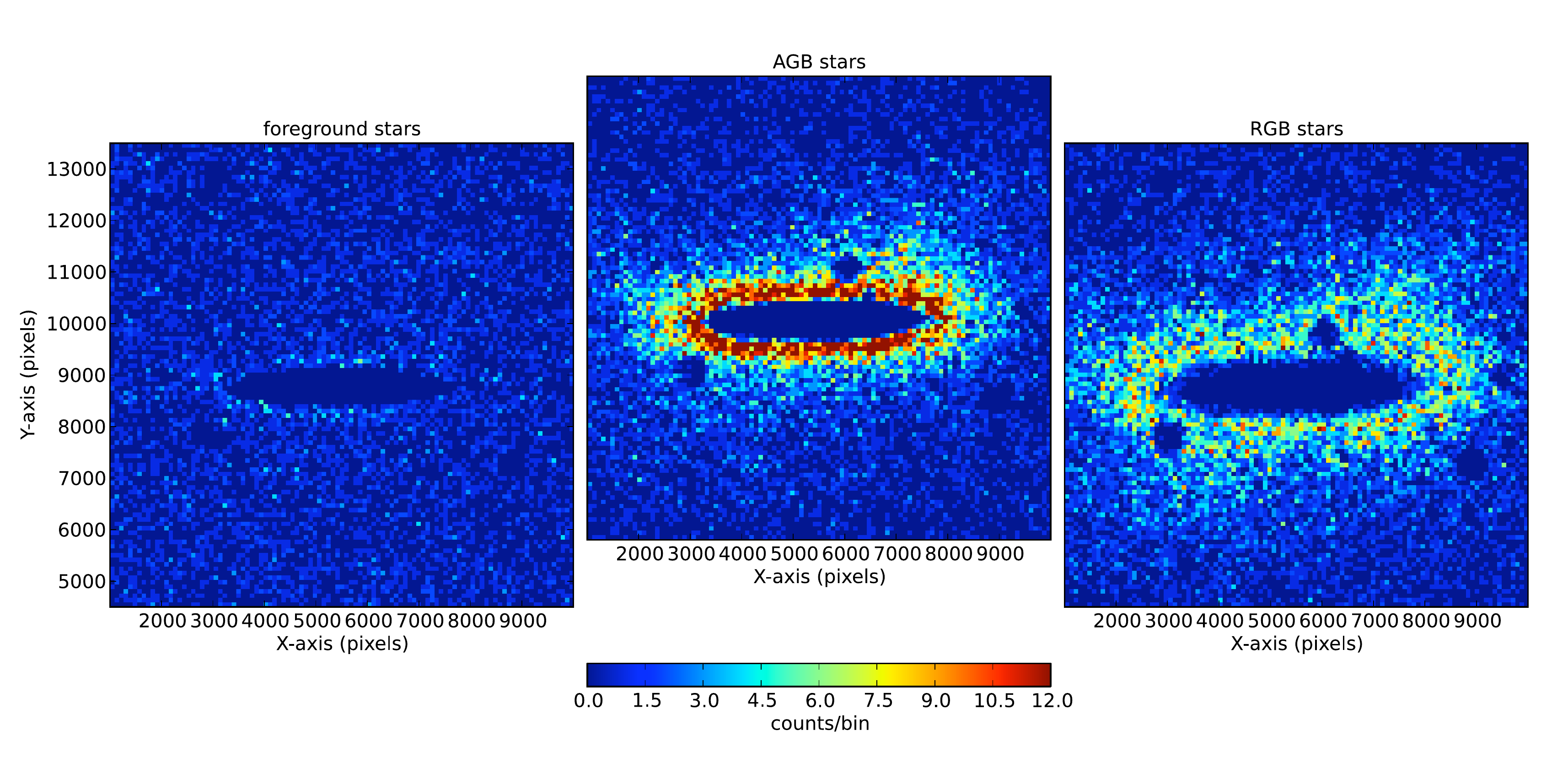}
}
\caption{The stellar density distribution of NGC\,253.   For each plot, the stellar catalogue has been binned to areas of 0.5 $\times$ 0.5 kpc (87.5 $\times$ 87.5 pixels)
and the density of each class of object computed.   The foreground population (left panel) shows a very uniform distribution with an average of 0.6 $\pm$ 0.8 stars
per bin.   The AGB stellar density (centre panel) steeply increases
toward the disk from background levels to a peak of 27 sources per bin.   In contrast, the RGB stars 
(right panel) are more uniformly distributed around the disk and extend further out.  The peak density of the RGB stars is 14 sources per bin.
}
\label{fig:AGBRGB_distr_disc}
\end{figure*}

The panoramic view of the stellar content of the halo of NGC~253 is
further illustrated on Fig.~\ref{fig:AGBRGB_distr_disc}, showing separately the
spatial density map of the foreground plus
background contamination, AGB and RGB stars. In the
central and right panels, the AGB and RGB averaged counts show the inner
structure of the halo and the off-planar perturbation of the disk. The
elongation corresponding to the southern shelf and its counterpart in
the north side are well visible, especially in the rightmost panel. It
is also apparent that the AGB stars are more concentrated in the disk
of the galaxy compared to the RGB component, although the different level of incompleteness affecting the two subpopulations is partly responsible for this effect. 
As already remarked,  some AGB stars are found out to large distances from the disk.

A crucial issue is the age of these AGB stars, and in general
the star formation history of the whole region sampled by VISTA. We
address this question by comparing the observed CMD to theoretical
isochrones. Very young stars are absent in the outer disk and halo of
NGC 253, and the most recent star formation episodes occurred about
0.3 - 0.5 Gyr ago.  The RGB stars are poor age indicators, but in our
CMD isochrones as old as 8 Gyr appear well populated. While we cannot
assess the age at which star formation began in NGC 253, the CMD is
consistent with the presence of stars as old as the oldest stars in
the Milky Way.  We estimate that the AGB population traces $\sim$ 2
$\times 10^{8}$ \msun\ of stars formed between 0.5 and 3 Gyr ago, with
75 $\%$ of this star formation occurring in the portion of the disk
sampled by our CMD. The remaining 25 $\%$ is spread out up to large
heights above the plane, $\simeq$ 30 kpc or more, as estimated from the
profile of the AGB stars along the minor axis (see Fig. \ref{fig:halo_extent}). 
The spatially resolved CMDs constructed for the
disk appear quite similar, except for subtle differences which need
appropriate simulations to be interpreted with confidence.

The presence of AGB stars over a wide volume as inferred from
Fig.~\ref{fig:AGBRGB_distr_disc}, is puzzling; either they were accreted, 
or stripped from/ejected out of the disk, or formed in situ. In the former two
cases, the event should have occurred a long time ago, so that the
accreted stellar population could mix well in the large volume
probed. This seems quite difficult to accomplish within a timescale of
1 Gyr, which is the typical age of the AGB stars. The third
possibility, i.e. in situ formation, could find support from the recent
detection of a massive molecular wind progressing at high speed from
the inner regions of NGC~253 \citep{Bolatto+13}, likely powered by the current
starburst. Although this presently ongoing outflow cannot be responsible for
the  AGB stars in our images, it shows that gas can reach large
distances in the galaxy halo. The north-west side of the wind appears
to be moving at $\sim 400$ km/sec. At this velocity the material would
reach a 30 kpc distance in 70 Myr. On the other hand, in order to form
stars, this gas must slow down and cool, which seems unlikely. In
addition, also in this case the smooth
distribution of the AGB component requires some substantial time delay
from formation to diffusion out of the birth place. Eventually, either
option, i.e. accretion or in situ formation, suffer from the same
problem, which is the need to mix the AGB population over a wide
volume within the stellar lifetime of $\sim$ 1 Gyr. As a possible
alternative, this population could be born from a diffuse, very large
cloud, but, in the absence of a nearby galaxy, no obvious triggering 
event is apparent.

\section{Conclusions}\label{sec:conclusions}

We present results based on deep $Z$ and $J$ band VISTA photometry
for a field centred on the spiral galaxy NGC~253, that extends up to
$\sim$ 40 kpc from the centre along the major axis (in the South-West
direction) and up to $\sim$ 50 kpc from the centre along the minor
axis (in the North-West direction).

We detect the smooth halo around NGC~253 and one 
substructure consisting of a $\sim 20$ kpc wide arc in the north west part
of the tile, at about 28~kpc above the plane. Assuming an average
metallicity of 0.008 and old age, the excess of 380 counts in this
substructure corresponds to a mass of the parent stellar
population of $\sim 7.5 \times 10^6 \msun$. 

The star counts suggest
that the halo consists of two components, an inner elliptical
structure superimposed on a more diffuse possibly spherical halo. Our
data are not deep enough to assess the shape of the external halo, but
the presence of member stars all over the tile, over more than 50~kpc
away from the disk is very well documented.

The wide view of this galaxy confirms the existence of an extra planar
region and the disturbed distribution of the outer disk, with the very
prominent \textit{southern shelf}, and a symmetrical feature 
on the north side. This argues for recent interaction for NGC~253,
possibly involving an important merging event with another galaxy.
However, the stellar population is remarkably
homogeneous in the outer disk, and virtually all over the tile. This
is surprising. There are strong indication of a recent merger, but no
young stars, no young clusters, and no obvious
substructures are evident. Particularly puzzling is the
distribution of AGB stars. Despite being more centrally concentrated
than the RGB stars, 25\% of the AGB population is spread all over the
halo out to about 30 kpc from the plane.  The spatial distribution and
origin of these stars, whose age should not be older than a few Gyr,
deserves further investigation.

\begin{acknowledgements}
We thank Jorge Melnick for initiating the VISTA Science Verification
and its organization, Thomas Szeifert and Monika Petr-Gotzens for the
assistance and help during the observing run, and Jim Lewis, Simon
Hodgkin and Eduardo Gonzalez-Solares from CASU for their expert
contribution to the VISTA data processing. We also thank Annette Ferguson 
and St\'ephane Courteau for interesting discussions, and the anonymous referee for several useful suggestions that improved the presentation of our results. L.G. thanks ESO for financial support and hospitality to work on this project.
\end{acknowledgements}

\bibliographystyle{aa}
\bibliography{N253_biblio}

\end{document}